\documentclass[floatfix,aps,prd,abstract,twocolumn,10pt,nofootinbib,preprintnumbers]{revtex4-1}

\usepackage{amsgen,amsmath,amstext,amsbsy,amsopn,amssymb}
\usepackage{graphicx}
\usepackage[usenames]{color}
\usepackage{epsfig}
\usepackage{multirow}
\usepackage{longtable}
\usepackage{hyperref}

\newcommand{\mb}{\mathbf}

\newcommand{\Ddel}{\delta_{\rm D}   }

\newcommand{\MpcOh}{ \,  \mathrm{Mpc}  \, h^{-1} }
\newcommand{\hOMpc}{ \,  \mathrm{Mpc}^{-1}  \, h  }

\newcommand{\comment}[1]{}
\newcommand{\nn}{ \nonumber }
\newcommand{\dv}{ \delta_{\rm v}}
\newcommand{\dc}{ \delta_{\rm c} }

\newcommand{\beq}{\begin{equation}}
\newcommand{\eeq}{\end{equation}}
\newcommand{\beqa}{\begin{eqnarray}}
\newcommand{\eeqa}{\end{eqnarray}}

\begin{document}


\title{ Large-Scale Clustering of Cosmic Voids  }
\author{Kwan Chuen Chan$^{(1)}$} \email{KwanChuen.Chan@unige.ch}
\author{Nico Hamaus$^{(2,3)} $}
\author{Vincent Desjacques$^{(1)}$} 
\affiliation{$^{(1)}$ D\'epartement de Physique Th\'eorique and Center for Astroparticle Physics,
Universit\'e de Gen\`eve, 24 quai Ernest Ansermet, CH--1211 Gen\`eve 4,
Switzerland}
\affiliation{$^{(2)}$ Sorbonne Universit\'es, UPMC Univ Paris 06, UMR 7095, Institut d'Astrophysique de Paris, F-75014, Paris, France}
\affiliation{$^{(3)}$ CNRS, UMR 7095, Institut d'Astrophysique de Paris, F-75014, Paris, France}

\date{\today}

\begin{abstract}
We study the clustering of voids using $N$-body simulations and simple theoretical models.  The excursion-set formalism describes fairly well the abundance of voids identified with the watershed algorithm, although  the void formation threshold required is quite different from the spherical collapse value.  The void cross bias $b_{\rm c} $  is measured and its large-scale value is found to be consistent with the peak background split results.  A simple fitting formula for  $b_{\rm c} $ is found.  We model the void auto-power spectrum taking into account the void biasing and exclusion effect. A good fit to the simulation data is obtained for voids with radii $\gtrsim 30\MpcOh$, especially when the void biasing model is extended to 1-loop order.  However,  the best-fit bias parameters do not agree well with the peak-background results.  Being able to fit the void auto-power spectrum is particularly important not only because it is the direct observable in galaxy surveys, but also our method enables us to treat the bias parameters as nuisance parameters, which are sensitive to the techniques used to identify voids.

\end{abstract}

\maketitle

\section{ Introduction}

Cosmic voids have emerged as an interesting probe of large-scale structure, as they account for the bulk of the cosmic web and can be easily observed in modern galaxy surveys. They are also good laboratories for testing general relativity \cite{Li2011,LiZhaoKoyama,ClampittCaiLi}, dark energy models \cite{ParkLee2007, LavauxWandelt_2010,  Biswasetal2012,Bosetal2012, LavauxWandelt_2012}, or inflationary non-Gaussianities \cite{KamionVoid,LamShethDesjacques2009,SongLee2009,DAmicoMussoetla2011} thanks to their low matter content. Moreover, voids preserve the initial conditions better than their overdense halo counterparts, as their evolution is simpler and has undergone less virialization \cite{Neyrinck_IC2012, NeyrinckYang2012}.

Although cosmic voids have been found in galaxy surveys for many years \cite{GregoryThompson,Kirshneretal1981}, little attention has been devoted to them as compared to halos. One of the reasons is that cosmic voids occupy large volumes while being the least sampled structures, so galaxy surveys are required to cover both large volumes and reach high sampling densities at the same time. Recent galaxy surveys like the SDSS have produced void catalogues suitable for statistical analyses \cite{Panetal2012,Sutter2012,NadathurHotchkiss2014,Sutteretal_2014}. Unfortunately, there is no well-defined definition for what is a void in galaxy surveys. Among the various possibilities (see \cite{Colbergetal2008} for a comparison of void finders), the void identification based on the watershed algorithm \cite{Platenetal2007} is a fairly general and practical definition because it does not require any prior on the morphology of voids and is parameter free.  We shall use it in this work. 

Most studies thus far have focused on the characteristics of individual voids, such as their average density profile \cite{ColbergShethetal2005,Padillaetal2005,Ricciardellietal2013, Pazetal2013, Ricciardellietal2014,HamausSutteretal2014,Nadathueetal2014}, but the large-scale spatial distribution and clustering of voids has been hardly investigated \cite{HamausWandeltetal2014,HamausSutterLavauxWandelt2014}. Voids have a much larger spatial extent than halos, ranging from a few to over 100 Mpc. Their size thus is expected to impact significantly their clustering statistics. In this paper, we shall study the clustering properties of voids, with the aim of eventually assessing how much cosmological information can be extracted from void measurements, and how much this technique is complementary to halo clustering statistics. For this purpose, we will model the void power spectrum using concepts similar to those developed for halo clustering studies. For example, the distribution of voids is also biased relative to the underlying dark matter distribution. However, void clustering statistics are much more sensitive to void exclusion effects owing to their large spatial extent. Understanding these effects will be essential to extract cosmological information from the large-scale clustering of voids.

The goal of this paper is to identify the most salient features in the large-scale clustering of voids using $N$-body simulations, and explain them with a simple model. The paper is organized as follows: we begin with a brief description of the simulations and the void finder used in this work (Sec.~\ref{sec:SimulationsFinder}). We examine the void size distribution and introduce the void bias parameters using the peak-background split (PBS) formalism (Sec.~\ref{sec:SizeDristributionPBSbias}). We then study the void cross-power spectrum, extract the void cross-bias parameter $b_{\rm c}$ and test whether it is consistent with the real-space void profile (Sec.~\ref{sec:VoidCrossPowerProfile}). We also measure the large-scale value of $b_{\rm c} $ and compare it with the PBS prediction.  Finally, we model the void auto-power spectrum in Sec.~\ref{sec:VoidPk_exclusion}, accounting for void exclusion using the hard-sphere model. Augmented with void biasing, we obtain a reasonably good description of the void auto-power spectrum for large voids. We discuss and conclude in Sec.~\ref{sec:conclusion}.

\section{Simulations and void finder}
\label{sec:SimulationsFinder}

Before presenting the numerical results, we first outline the details of the simulations used in this paper. Two suites of simulations with different box sizes are used: 1500 $\MpcOh $ (six realizations) and 250 $\MpcOh $ (three realizations), abbreviated as L1500 and L250 later on. In each simulation, there are $1024^3$ particles. The cosmology is a flat $\Lambda$CDM model, with the WMAP~7 cosmological parameters adopted \cite{WMAP7}, i.e.~, $\Omega_{\rm m} = 0.272$, $\Omega_{\Lambda}=0.728$, $\Omega_{\rm b} = 0.0455$, and  $\sigma_8=0.81$. Thus, for the large box each particle carries a mass of $2.37 \times 10^{11} \, M_{\odot} h^{-1} $, while this value is $1.10 \times 10^{9} \, M_{\odot} h^{-1} $ for the small box. The combination of large and small box sizes enables us to capture a wide range in void sizes and to conduct a resolution study. We use Gaussian initial conditions with a spectral index of $n_s=0.967$. The transfer function is output from {\small CAMB} \cite{CAMB} at redshift $z=99$. The initial particle displacements are implemented using {\small 2LPT} \cite{CroccePeublasetal2006} and the simulation is run with the {\small Gadget2} code \cite{Gadget2}. Furthermore, although we do not show them explicitly here, many of the results presented have been cross-checked using the $1000 \MpcOh $ simulation used in \cite{HamausWandeltetal2014,HamausSutteretal2014}.

Voids are identified using the public void identification and examination toolkit {\small VIDE} \cite{Sutteretal2014}. At its core, the void finder {\small ZOBOV} \cite{Neyrinck2008}, which is based on the watershed algorithm \cite{Platenetal2007}, is employed. In this work, we will only consider voids identified in the dark matter distribution.  Since many of their properties are sensitive to the exact void finding procedure, we begin with a brief description of our void identification algorithm. We refer the reader to \cite{Neyrinck2008,Sutteretal2014} for more details. First, tracer particles are partitioned by a Voronoi tessellation, and a density field is created. The Voronoi cells around a local minimum are joined to form catchment basins (zones). For each zone, the density (water level) is increased until another local minimum is found (water flows to a new local minimum) and the height of the ridge separating the two minima (watershed) is recorded. 
We stop growing voids if the ridge between zones is higher than $0.2\bar{n} $, where $\bar{n} $ is the mean particle number density \cite{Neyrinck2008,Sutteretal2014}. However, we note that this lead to asymmetry in the treatment between the voids with  minimal density less then $0.2\bar{n} $, which are allowed to grow the void hierarchy, and those voids with  minimal density larger than $0.2\bar{n} $, which can only be single zone voids. This asymmetry only exists for the smallest voids, because the minimum densities of larger voids are always below $0.2\bar{n} $. 
Apart from this criterion, there are no other free parameters in the void finder itself, and generally voids can take on any shape. However, one can further apply various cuts on the catalog to select some specific samples.  In the void finder, there is no a priori upper bound for their minimum density.  The void volume $V$ is calculated as the sum of the void's constituent Voronoi cell volumes. The void size or effective radius is given by 
\beq
R = \Big( \frac{ 3 V }{ 4 \pi  } \Big)^{ \frac{1}{ 3}  }, 
\eeq
and we define the void center as volume-weighted average of all its Voronoi cells,
\begin{equation}
  {\bf X} = \frac{1}{\sum_i V_i} \sum_i {\bf x}_i V_i,
\label{eq:macrocenter}
\end{equation}
where ${\bf x}_i$ and $V_i$ are the positions and Voronoi volumes of each tracer particle $i$, respectively. In this paper, we consider voids identified in dark matter simulations only.  We note that there are further complications in going from dark matter voids to  galaxy voids, but these are fairly well understood \cite{FurlanettoPiran_2006,Sutteretal2013,Sutteretal2_2014}.

The watershed algorithm automatically returns a hierarchical structure for parent voids and their nested children (sub-voids). Most of the existing void studies focus on parent voids only.  The sub-void fraction in a given sample increases with the tracer sampling density, but as voids exhibit self-similar behavior \cite{Sutteretal2013}, sub-voids share many properties of the parent voids \cite{HamausSutteretal2014}. Thus, the inclusion of sub-voids increases the sample size and therefore its statistical power. In this work, we will only  show the results obtained using all the voids returned by {\small ZOBOV} that are larger than the mean particle separation, 
without additional filtering. However, we have also checked the robustness of our results using two distinct sub-samples: one including the parent voids solely, and the other including the voids with central density $\leq 0.2 \bar{n}  $. The central density is defined to be the one averaged within $R/4 $ about their center.  However, we note that this kind of cut can be noisy for small voids because of few number of particles inside $R/4$, but it should be fine for large voids.  Also, the central density defined here should not be confused with the minimal density at the center of the voids.  We will comment on the differences in the results from various samples in due course.

The tracer sampling density is a key quantity of the void finding process. In order to study its influence on void properties, we randomly remove tracer particles from our simulations to achieve different degrees of sub-sampling. For L1500, we use 0.02 $ (\MpcOh)^{-3} $, which roughly amounts to the number density of galaxies in the lowest redshift main sample of the SDSS. We use three different sampling densities for L250, namely 2, 0.2 and 0.02  $ (\MpcOh)^{-3} $. For the relatively low sampling density of 0.02 $( \MpcOh )^{-3}$, the sample is dominated by top-level voids, while for the sample with 2 $( \MpcOh )^{-3}$ tracer density, the contribution from sub-voids is much more important.  The hard sphere model that we consider is no longer effective when sub-voids are included, thus we need to discard them in that case.

\section{Size distributions and the Peak-Background split bias}
\label{sec:SizeDristributionPBSbias}

In this section, we shall present the numerical void size distribution and compare it with the theoretical one obtained from the excursion-set formalism. We will also derive the peak-background split (PBS) bias parameters. 

\subsection{ The void size distribution} 
Like halos, the abundance of voids can be modeled using the excursion-set formalism \cite{BCEK1991}. It naturally solves the so-called cloud-in-cloud problem, which arises when a halo is embedded in a larger-scale overdensity and subsumed in the latter to form an even bigger halo. On the contrary, there is no cloud-in-void problem, because parent halos can exist in large-scale underdensities. Therefore, there is only one barrier in the excursion set description of halos, which is the linear threshold $\dc $ for halo formation. For the description of voids however, Sheth and van de Weygaert (SvdW) \cite{ShethWeygaert} argued that one needs to consider both $\dc $ and the threshold for void formation $\dv $. Besides the void-in-void problem (analog to the could-in-cloud problem of halos), voids are subject to the void-in-cloud problem: a void embedded within a collapsing overdensity will be crushed out of existence. SvdW argued the value of $\dc$ lies somewhere between the linearly extrapolated overdensity at turn-around and at virialization, or, according to the spherical collapse model, between 1.06 and 1.68, respectively.  The epoch of shell crossing is used to define the threshold for void formation $\dv $ \cite{Blumenthaletal_1992}, which is about $-2.81$ according to spherical expansion in the Einstein-de Sitter Universe.

In analogy with the halo mass function, the void mass function can be cast into the form
\beq
\frac{ d n }{ d \ln M } = \frac{ \bar{ \rho}_{\rm m}  }{ M } \nu \mathcal{F}( \nu, \delta_{\rm v},  \delta_{\rm c}  )  \frac{ d \ln \nu } {  d \ln M }, 
\eeq
where $\bar{\rho}_{\rm m} $ is the mean dark matter density and 
\beq
\nu = \frac{ | \delta_{\rm v} | }{ \sigma_M  } 
\eeq
is the peak height or significance.
Here, $ \sigma_M $ is the root-mean-squared density fluctuation smoothed with a top-hat window of size $R_{\rm L}$, the Lagrangian size of the void. The first crossing distribution $\mathcal{F} ( \nu, \delta_{\rm v},  \delta_{\rm c}  ) $  denotes the probability that a random trajectory first crosses the barrier $  \delta_{\rm v} $ at $ \nu $ without crossing $ \delta_{\rm c} $ for $\nu' > \nu $. It is given by  \cite{ShethWeygaert}
\beq
\label{eq:Fvoid_nu}
\mathcal{F}  ( \nu ) 
=  \frac{ 2 \mathcal{D}^2 }{\nu^3 } \sum_{j=1}^{\infty}   j \pi   \sin (\mathcal{D} j \pi)  \exp \Big(  - \frac{j^2 \pi^2 \mathcal{D}^2 }{ 2  \nu^2  }     \Big) , 
\eeq
where $\mathcal{D}$ is the void-and-cloud parameter
\beq
\label{eq:VoidCloudpara}
\mathcal{D} = \frac{  - \delta_{\rm v}   }{ \delta_{\rm c} -  \delta_{\rm v} }.
\eeq
SvdW provided an approximate, albeit more compact expression for the series in Eq.~\ref{eq:Fvoid_nu}:
\beq
\label{eq:Fapprox}
 \mathcal{F}_{\rm approx}  ( \nu ) 
= \sqrt{ \frac{2  }{ \pi } } \exp \Big( - \frac{ \nu^2 }{ 2 }  \Big)  \exp\Big(  - \frac{| \delta_{\rm v} | }{\delta_{\rm c}  } \frac{ \mathcal{D}^2 }{ 4 \nu^2 }  - 2 \frac{\mathcal{D}^4  }{\nu^4}  \Big).
\eeq

For large $M$, the Lagrangian void contains a large amount of matter, such that the first-crossing distribution is dominated by random walks that first cross $\dv $ directly, without reaching relatively large positive $ \delta $. In this regime, $\mathcal{F}(\nu)$ reduces to the probability of crossing only one barrier, $ \dv $:  
\beq
\label{eq:OneBarrier}
\mathcal{F}_{\rm one}  ( \nu ) = \sqrt{ \frac{2  }{ \pi } } \exp \Big( - \frac{ \nu^2 }{ 2 }  \Big).  
\eeq
In fact, most of the voids we consider in this paper fall into this regime. 
  

In observational data, the relevant quantity is the void size distribution rather than the void mass function. The simplest way to convert the mass in Lagrangian space to size in Eulerian space is to use the spherical collapse model. In spherical collapse, the nonlinear density contrast of voids is $-0.8$. From this, we can estimate the Lagrangian size of voids, $R_{\rm L} $:
\beq
\label{eq:RL_R_relation}
R_{\rm L} = 0.58 R.
\eeq
Assuming their number density is conserved when voids evolve from Lagrangian to Eulerian space, the void size distribution in Eulerian space becomes 
\beq
\frac{d n }{d \ln R } =  \frac{ d n }{ d \ln R_{\rm L} } = 3 \frac{d n   }{ d \ln M },
\eeq
where we have used Eq.~\ref{eq:RL_R_relation} to map $R$ to $R_{\rm L} $.

\begin{figure}[!htb]
\centering
\includegraphics[width=\linewidth]{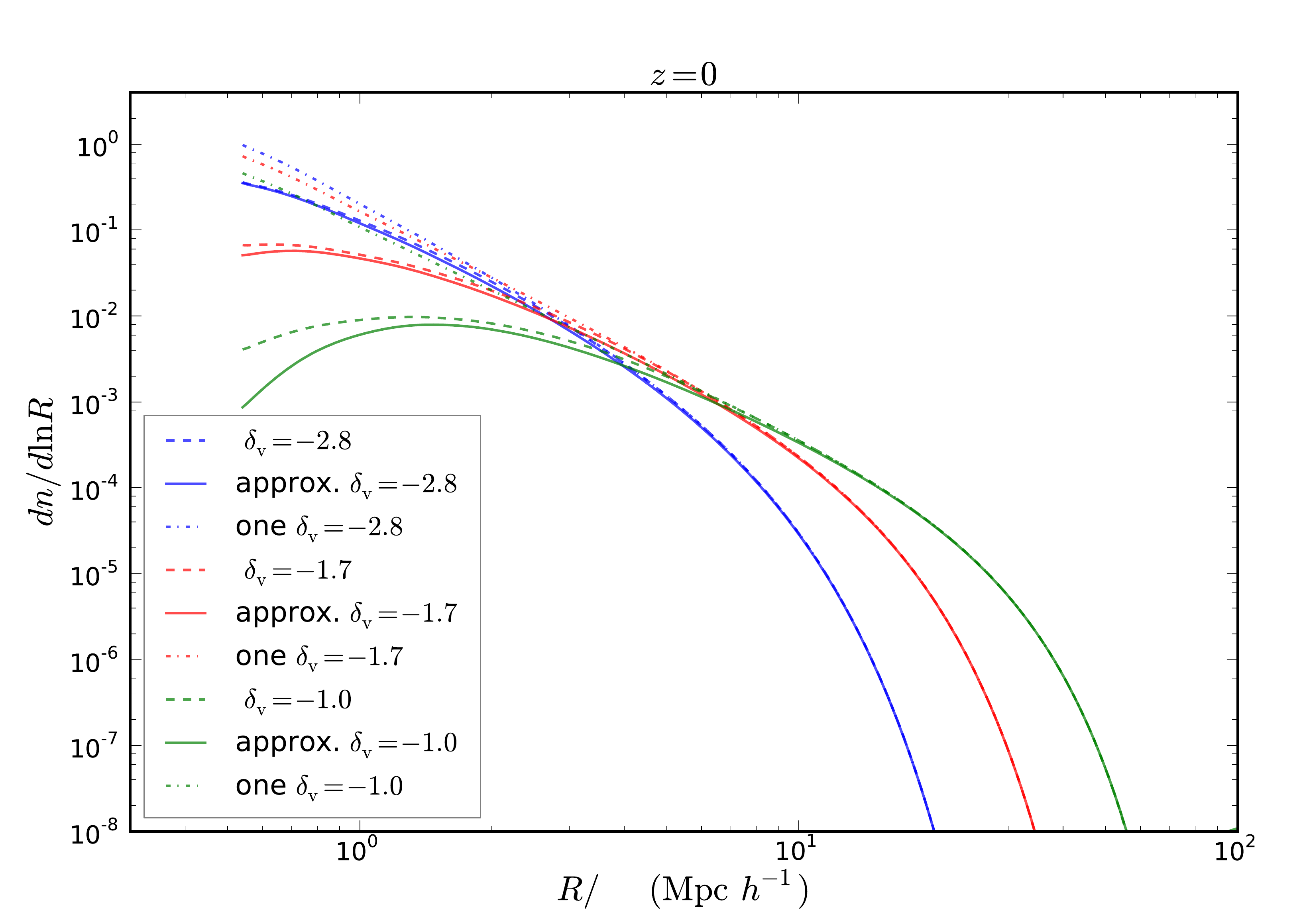}
\caption{  The void size distribution for various values of $\dv$: $-2.8$ (blue),   $-1.7$ (red) and   $-1.0$ (green). For each value, the void size distributions obtained using Eq.~\ref{eq:Fvoid_nu} by summing up the first 12 terms (dashed), the approximate distribution from Eq.~\ref{eq:Fapprox} (solid), and the single barrier distribution from Eq.~\ref{eq:OneBarrier} (dotted-dashed) are compared. }
\label{fig:void_size_dist_SvdW}
\end{figure}

Since the void abundance sensitively depends on the void definition, we shall treat $\dv$ as a free parameter to model the data. On the other hand, for the large voids that we are mainly interested in, the effect of $\delta_{\rm c}$ is negligible, unless its value is much smaller than the spherical collapse threshold of 1.68. Thus, we will simply fix $\dc$ to be 1.68 throughout. For illustration, Fig.~\ref{fig:void_size_dist_SvdW} displays the void size distribution for various values of $\dv$ using the exact SvdW first crossing distribution Eq.~\ref{eq:Fvoid_nu}, where we sum up to the first 12 terms of the series. The approximate distribution from Eq.~\ref{eq:Fapprox} is also shown. It provides a very good agreement to Eq.~\ref{eq:Fvoid_nu} for $R\gtrsim 5 \MpcOh $. In this regime, even the one-barrier distribution Eq.~\ref{eq:OneBarrier} is a good approximation to the exact result. 

\begin{figure*}[!htb]
\centering
\includegraphics[width=\linewidth]{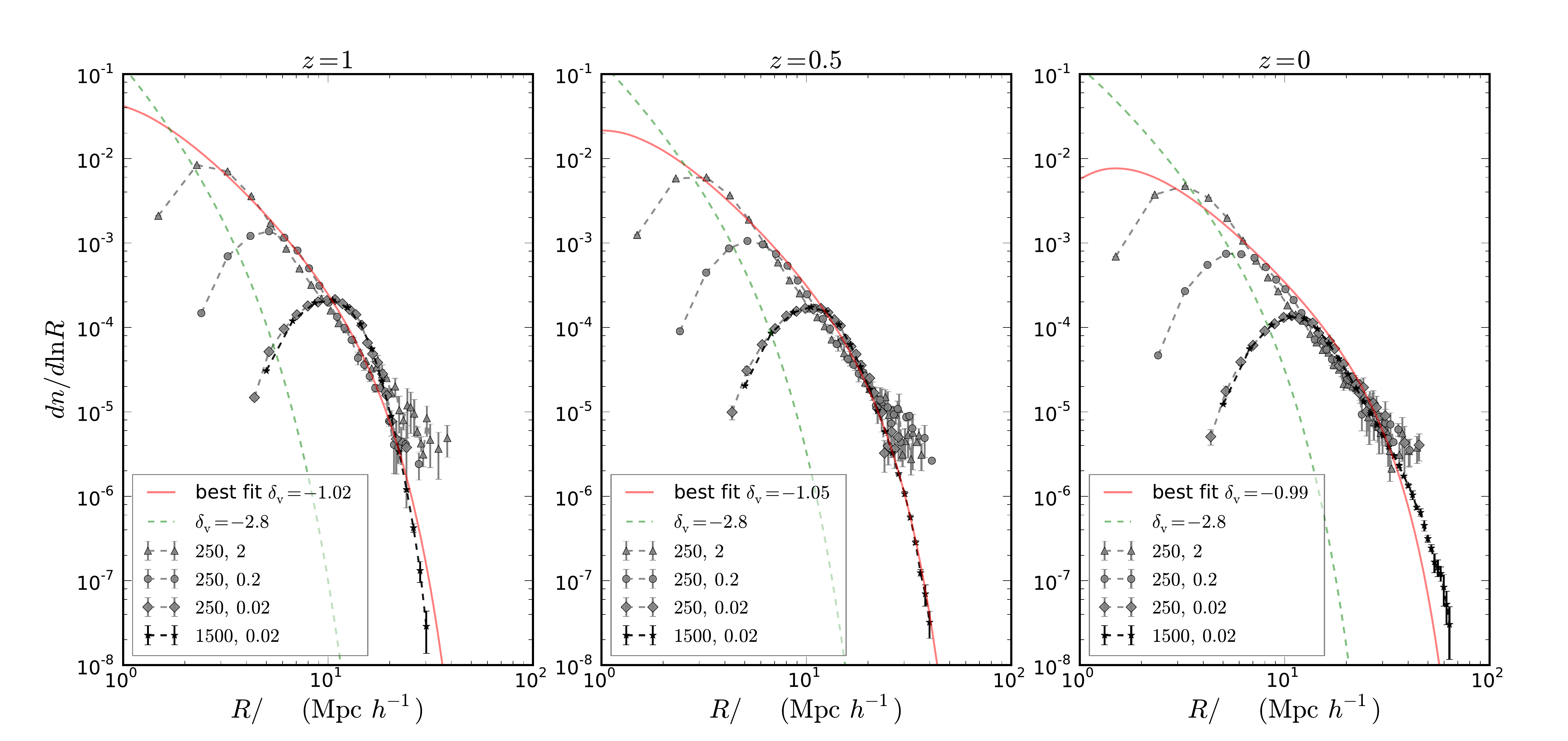}
\caption{  The void size distribution  at $z=1$, 0.5 and 0 (from left to right) measured in simulations.  The data are obtained from L1500 with sampling density 0.02 $(\MpcOh)^{-3}  $ (star, black), and  L250  with sampling densities 2 (triangle, gray), 0.2 (circle, gray), and 0.02  $(\MpcOh)^{-3}  $ (diamond, gray). The SvdW void size distribution is shown with $\delta_{\rm v}$ equal to $-2.8$ (dashed, green) and with the best-fit value $\dv  $ (solid, red) as stated in the inset of each panel. }
\label{fig:void_size_dist}
\end{figure*}

The void size distributions measured in our $N$-body simulations at $z=1$, 0.5 and 0 are depicted in Fig.~\ref{fig:void_size_dist}. Note that, for a sampling density of 0.02 $(\MpcOh )^{-3}$, both L1500 and L250 agree nicely on the abundance of small voids. However, the L1500 simulation samples the large voids much better, and thus extends the void size distribution to larger void radii. To obtain measurements at small void radii, we increase the sampling density in L250. While the L1500 catalogues are dominated by top-level voids, the smaller boxes with larger sampling density are more influenced by sub-voids. Notwithstanding, the data points in Fig.~\ref{fig:void_size_dist} agree reasonably well. However, there is some discrepancy between different sampling densities in the overlapping regions: higher sampling densities seem to suggest lower abundances for intermediate-size voids. This effect can be explained by void fragmentation due to resolution effects~\cite{Sutteretal2013}: only with sufficient sampling density can the sub-structure of all voids be resolved. If this is not the case, small voids artificially merge to form larger voids.


In Fig.~\ref{fig:void_size_dist}, we overplot the SvdW prediction with $ \dv=-2.8 $ together with the best-fit void size distribution obtained upon allowing $\dv $ to vary freely (as in \cite{Sutteretal2013}). As the higher sampling densities are more affected by sub-voids, and sub-voids are not included in the treatment of SvdW, we only fit the L1500 data with $R>20 \MpcOh $. We fit the data separately at each redshift and obtain the best-fit values of $\dv= -1.02$, $ -1.05$ and $-0.99$ for $z=1$, 0.5 and 0, respectively. These numbers are remarkably consistent with each other, although they are quite different from the canonical spherical collapse value of $-2.8$.  In the construction of voids, the shell crossing condition, which is used to define void formation,  is never incorporated. A priori it is not clear if the voids constructed in the watershed algorithm would agree with the shell crossing estimate.  Because the watershed algorithm defines voids of arbitrary geometries and density profiles in various tracer sampling densities, this discrepancy is not surprising.

Interestingly, the best-fit curves also  agree with the void size distributions from L250 down to lower void radii. In fact, if we restrict ourselves to top-level voids only, the void abundance hardly changes for a sampling density of 0.02 $(\MpcOh )^{-3}$. 
At higher sampling densities however, it is reduced by a factor of at least a few and does not agree with the best fit anymore.



There are some extensions of the SvdW model which address mainly the void-in-cloud problem \cite{ ParanjapeLamSheth2012, AchitouvNeyrinckParanjape2013, Jenningsetall2013}. As we are mostly interested in the larger voids, these are not relevant here, so we restrict ourselves to the simpler SvdW model.  In the original SvdW model, the steps in the excursion set are uncorrelated, which corresponds to a sharp-$k$ filter. In \cite{ParanjapeLamSheth2012,AchitouvNeyrinckParanjape2013}, the excursion set with correlated steps was considered. However, correlated steps reduce the void abundance \cite{ParanjapeLamSheth2012} and, therefore, do not improve the agreement with our numerical data. In the SvdW model, the void number density $dn$ is conserved as one converts the void size distribution from Lagrangian to Eulerian space. In Ref.~\cite{Jenningsetall2013} the authors argued instead that large voids form by merging of smaller voids and, thereby, the volume $Vdn $ should be conserved. Using this $Vdn$ model, together with Eq.(\ref{eq:RL_R_relation}), one obtains
\beq
\frac{d n }{d \ln R }  =  \frac{  V_{\rm L } }{ V }  \frac{d n_{\rm L} }{d \ln R_{\rm L} }, 
\eeq
where the Lagrangian quantities are denoted by the subscript L, while the Eulerian ones are without subscript.  Ref.~\cite{Jenningsetall2013} found that this prescription results in much better agreement with simulation data than the original SvdW model. Because in spherical collapse, ${V } > {V_{\rm L}  } $, this model with $\delta_{\rm v}=-2.8 $ yields a worse fit to our simulation data than SvdW. Although \cite{Jenningsetall2013} also use {\small ZOBOV} to identify voids, they only select spherical non-overlapping regions of density $0.2 \bar{n}$ around density minima as voids, whereas we include all voids returned by {\small ZOBOV} without any further processing. Also note that Ref.~\cite{Jenningsetall2013} focus on voids with radii $R\lesssim 20 \MpcOh$.  This reflects the important caveat that the void size distribution is very sensitive to the void definition adopted.


\subsection{Void bias from the peak-background split }

As demonstrated in the previous section, by allowing $ \dv  $ to vary as a free parameter in the excursion-set formalism, we can fit the void size distribution reasonably well. We can now use the void mass function to derive the PBS bias parameters for voids. For $ R_{\rm v} \gtrsim 5 \MpcOh $ the full SvdW mass function is well approximated by Eq.(\ref{eq:Fapprox}). As we are mostly interested in voids of that size range, we will use this approximation for convenience.

Suppose there is a long wavelength perturbation $\delta_{\rm L} $ in the Lagrangian space, then the thresholds $\delta_{\rm v}   $ and  $\delta_{\rm c}   $ shift as 
\beq
\delta_{\rm v} \rightarrow \delta_{\rm v} - \delta_{\rm L}  , \quad    \delta_{\rm c} \rightarrow \delta_{\rm c} - \delta_{\rm L}.
\eeq
The bias parameters in Eulerian space are given by (see e.g. \cite{MyThesis})
\beq
b_i = \frac{1}{ n_0 } \frac{ \partial^i   } { \partial \delta^i }   [ ( 1 + \delta  ) n( \delta_{\rm L} )  ] \Big|_{  \delta = 0} ,  
\eeq
where $\delta  $ denotes the corresponding perturbation in Eulerian space and  $n_0  $ and  $ n( \delta_{\rm L} )$  represent the mass function with zero and $\delta_{\rm L}  $ background perturbation, respectively.  The factor $1 + \delta$  maps the mass function from Lagrangian space to Eulerian space. We shall use spherical collapse to relate $\delta $ and $\delta_{\rm L} $ \cite{Bernardeau1994},
\beq
\delta_{\rm L} = \delta - \nu_2 \delta^2 + \nu_3 \delta^3  + \dots,
\eeq
with $\nu_2 = 12/21  $ and  $\nu_3 = 341/567  $. Therefore, the bias parameters are given by
\begin{widetext}
\beqa
\label{eq:b1void}
b_1 &=& 1 + \frac{ \nu^2 -1 }{ \delta_{\rm v}  } + \frac{ \delta_{\rm v} \mathcal{D}  }{ 4 \delta_{\rm c}^2  \nu^2  },               \\
\label{eq:b2void}
b_2 &=&  \frac{ 2( \nu_2 -1 ) }{ \delta_{ \rm v} } + \frac{ \mathcal{D} }{ 2 \delta_{\rm c}^2  } + \frac{ \nu^2  } {\delta_{\rm v}^2   } [  2 \delta_{\rm v} ( 1 - \nu_2 ) - 3   ]  + \frac{ \nu^4 }{ \delta_{\rm v}^2  }  \frac{ \mathcal{D} \delta_{\rm v} }{ 2 \delta_{\rm c}^2 \nu^2  } \Big(   1 - \nu_2 + \frac{ 1 }{ \mathcal{D} \delta_{\rm c} }           \Big) + \frac{\mathcal{D}^2 \dv^2 }{ 16  \dc^4 \nu^4  }  , 
\eeqa
and 
\beq
\label{eq:b3void}
b_3 = 3 [ b_2 - 2 ( b_1 -1 ) ]  + b_3',
\eeq
where 
\beqa
b_3' & =&  \frac{\nu^6  }{ \dv^3 }  - \frac{ 6( 1 + \nu_2 \dv )   }{ \dv^3  } \nu^4  - \frac{ 3 \mathcal{D} \nu^2  }{  4 \dc^2 \dv^4 }    \big[  -\dv^3 + 4 \dc^3 ( 1 + 6 \nu_2 \dv + 2 \nu_3 \dv^2 )  -  4 \dc^2 \dv ( 1 + 6 \dv \nu_2 + 2 \dv^2 \nu_3   )   \big]  \nn \\
&+& \frac{ 3 \mathcal{D}^2  }{ 16 \dc^4 \dv \nu^2  } [ -9 \dv^2 + 16 \dc \dv (  1 + \dv \nu_2) + 8 \dc^3 ( 2 \nu_2 + \dv \nu_3 ) - 8 \dc^2 ( 1 + 4 \dv \nu_2 + \dv^2 \nu_3  )     ]  \nn \\
&-& \frac{3 \mathcal{D}^2 \dv  }{ 16 \dc^5 \nu^4 } ( \dc - 2 \dv + 2 \dc \dv \nu_2   )
+ \frac{ \mathcal{D}^3 \dv^3  }{ 64 \dc^6 \nu^6  }.
\eeqa
\end{widetext}
Eq.~\ref{eq:b1void} corrects a typo in Eq.~27 in Ref.~\cite{ShethWeygaert}. When $\mathcal{D}=0$, Eq.~\ref{eq:Fapprox} reduces to Eq.~\ref{eq:OneBarrier}. Thus, Eqs.~\ref{eq:b1void}$-$\ref{eq:b3void} reduce to those obtained from the Press-Schechter mass function \cite{MoWhite, MoJingWhite}.

\begin{figure}[!htb]
\centering
\includegraphics[width=\linewidth]{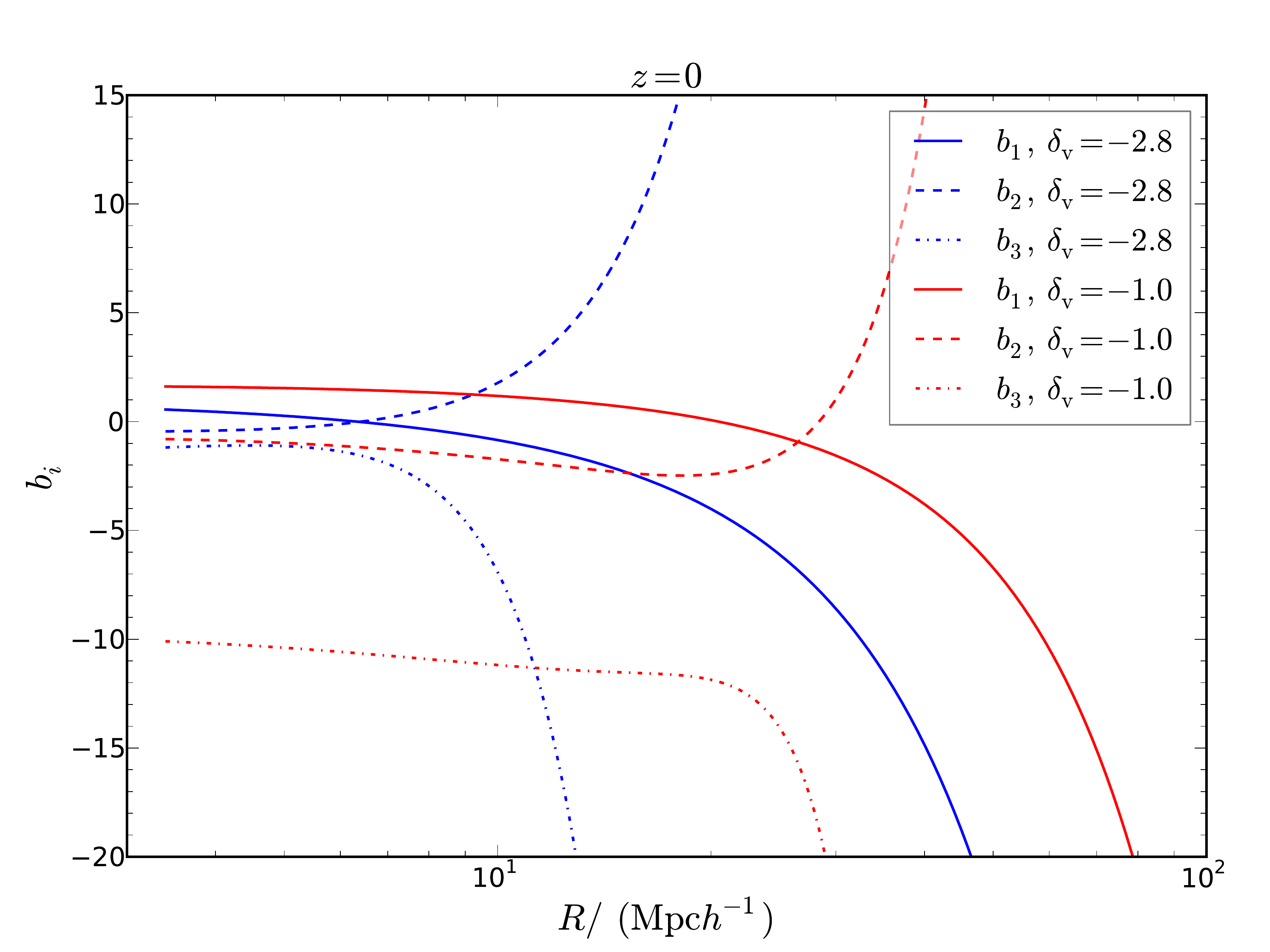}
\caption{  The PBS bias parameter $b_1 $ (solid), $b_2$ (dashed) and $b_3$ (dotted-dashed) as a function of void radius $R $. Two values of $\dv$ are chosen: $-2.8$ (blue) and $-1.0$ (red).  }
\label{fig:PBSbias_REul}
\end{figure}

In Fig.~\ref{fig:PBSbias_REul} we show $b_1$, $b_2$, and $b_3$ as a function of $R$, each for the two values of $ \dv =-2.8$ and $-1.0$ at redshift $z=0$.  Notice that $b_1$ and $b_2$ cross zero at similar values of $R$. In the following Sec.~\ref{sec:VoidCrossPowerProfile} and \ref{sec:VoidPk_exclusion}, we measure the bias parameters from the cross-power spectrum and the auto-power spectrum of voids individually, and compare them with the PBS results.

\section{Cross-power spectra and density profiles } 
\label{sec:VoidCrossPowerProfile}

In this section, we measure the cross-power spectrum between voids and the dark matter in our simulations and extract the large-scale bias parameters from it. We also determine the void density profile in configuration space, which is the Fourier transform of the void-matter cross-power spectrum. Since in practice small inaccuracies and noise either in configuration or Fourier space may significantly affect the corresponding transform, we find it prudent to investigate the correspondence between the two measurements in the data.

\subsection{The void cross-power spectrum }

The void density contrast $\delta_{v} $  is defined in analogy to that of halos, i.e.
\beq
\delta_{v} \equiv \frac{ n_{\rm v} - \bar{n}_{\rm v}  }{  \bar{n}_{\rm v} }, 
\eeq
where $ n_{\rm v}$ and $ \bar{n}_{\rm v} $ are the number density and the mean number density of voids, respectively. The cross-power spectrum $P_{\rm c }$ between voids and dark matter is defined as 
\beq
\langle \delta_{v} (\mb{k}_1 )   \delta  (\mb{k}_2 )  \rangle  = P_{\rm c } ( k_1 ) \Ddel( \mb{k}_1 + \mb{k}_2 ), 
\eeq
where $\delta$ is the dark matter density contrast and $\Ddel$ the Dirac delta function.  Using $P_{\rm c} $, we define the cross bias parameter as
\beq
b_{\rm c} \equiv  \frac{ P_{\rm c}}{ P_{\rm m} },
\eeq
where $P_{\rm m}$ denotes the dark matter auto-power spectrum. 

\begin{figure*}[!htb]
\centering
\includegraphics[width=\linewidth]{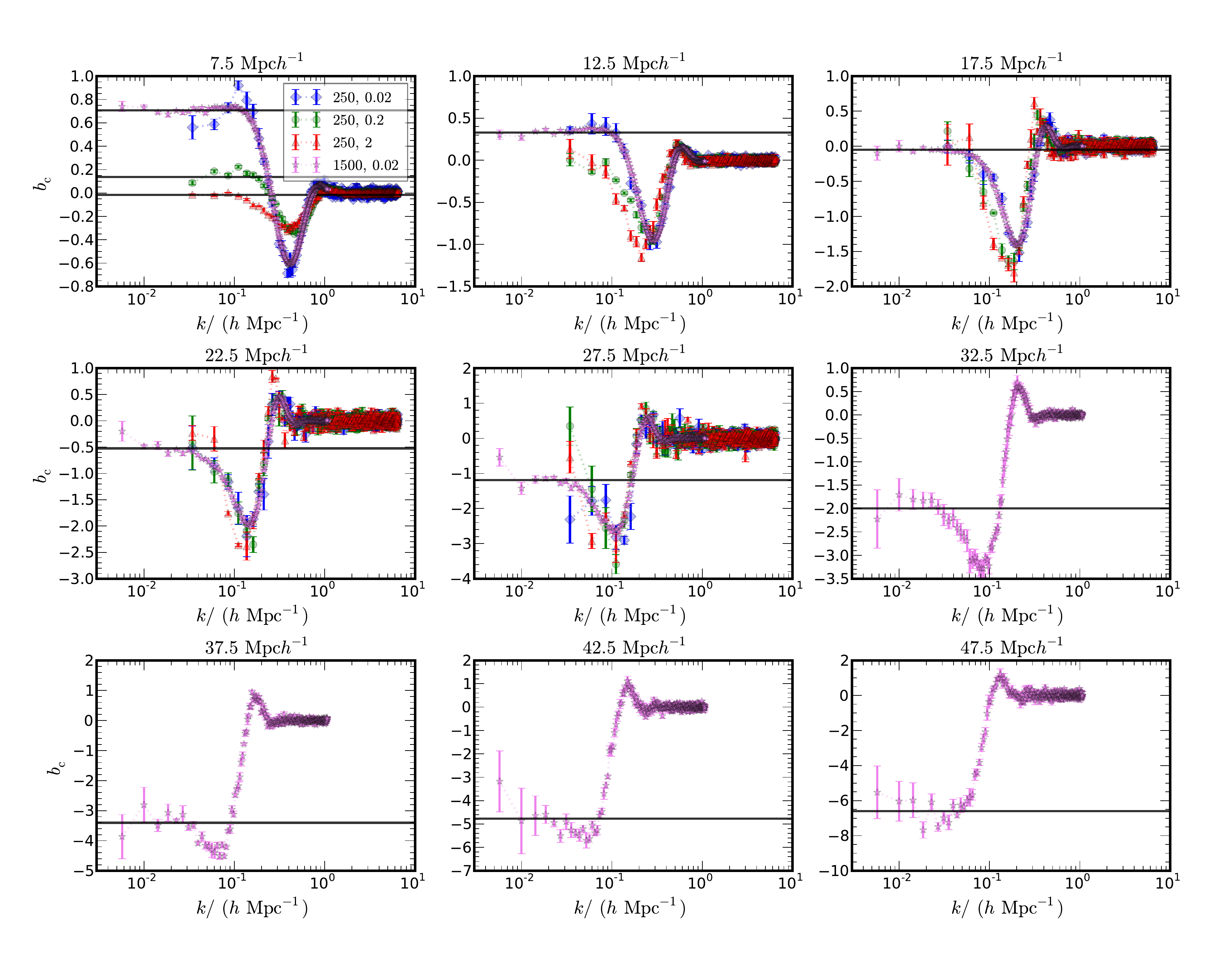}
\caption{ Void-matter cross-bias $b_{\rm c}$ as a function of wavenumber $k$ for various void sizes $R$ at $z=0$. The results are shown for L1500 with sampling density 0.02 $(\MpcOh)^{-3}$ (star, purple), L250 with sampling density 0.02 (diamond, blue), 0.2 (circle, green), and 2 $(\MpcOh)^{-3}$ (triangle, red). Horizontal lines show best fits to the linear large-scale regime of $b_{\rm c}$ whenever the fit is feasible (solid, black). } 
\label{fig:bc_z0_BoxCompare}
\end{figure*}

In Fig.~\ref{fig:bc_z0_BoxCompare} we plot $b_{\rm c}(k)$  at  $z$=0.  We have grouped all voids into radius bins of width 5 $\MpcOh $, the mid-points of the bin values are stated on top of each panel. 
For  $ R  \lesssim 20 \MpcOh  $, the behavior of  $b_{\rm c}$ at low $k$ changes with the sampling density. The cross bias of the larger voids ($ R \gtrsim  20  \MpcOh $) is more robust to variations in the sampling density, yet the results are noisier owing to sampling variance, especially in the small box. Despite the strong dependence of small voids on sampling density, the existence of significant correlations (and anti-correlations) between voids and the dark matter suggests the possibility of extracting information on dark matter from void clustering.


Overall, our results agree with the findings of Ref.~\cite{HamausWandeltetal2014}: $b_{\rm c}$ is roughly constant at low $k$ ($ k \lesssim 1/ R$), whereas it exhibits oscillations on intermediate scales and converges to zero at high $k$ ($k\gtrsim 10/R$). The relative scaling is more easily recognizable when $b_c$ is plotted against $ kR $, see Fig.~\ref{fig:bck_k4GaussFit_z0}. The qualitative features also hold for other redshifts, such as  $z=1$.  Note that the radius of voids with $b_{\rm c}(k\to0)=0$ corresponds to about $17.5\MpcOh$, in agreement with the number Ref.~\cite{HamausSutteretal2014} found for the size of compensated voids in the dark matter. Because compensated voids do not feature any large-scale perturbations, their linear bias vanishes~\cite{HamausWandeltetal2014}.    
In the following section, we will show how the oscillations in $b_{\rm c}$ are related to the structure of the void density profile.

\begin{figure*}[!htb]
\centering
\includegraphics[width=\linewidth]{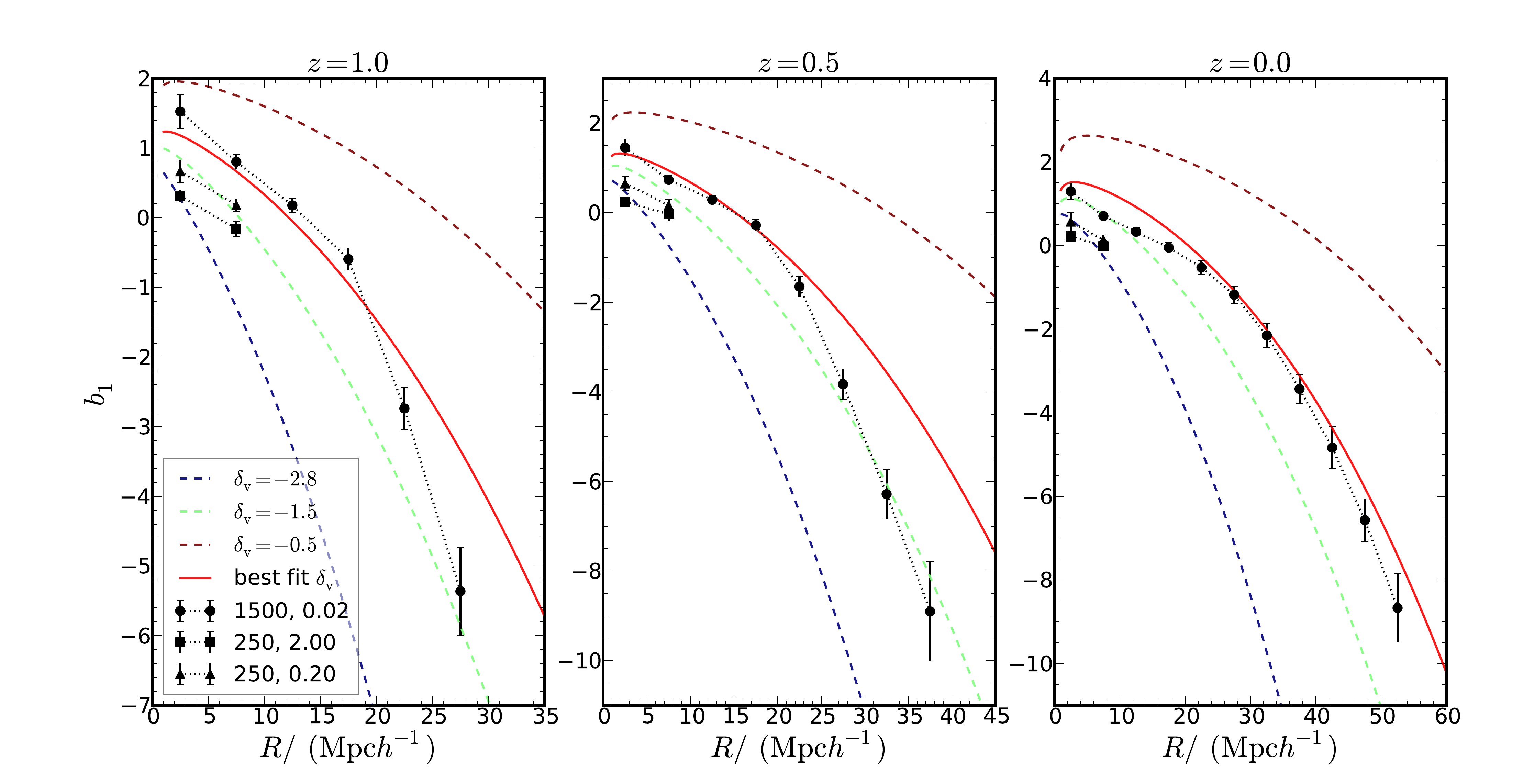}
\caption{  The large-scale best fit to the void-matter cross-bias $b_{\rm c}$ as a function of void radius $R $ from L1500 (circles), L250 with sampling density 2 (square) and 0.2 $( \MpcOh)^{-3}  $ (triangle) at redshifts $ z=1 $, 0.5 and 0 (from left to right). The curves show $b_1 $ computed from Eq.~\ref{eq:b1void} for various values of $ \dv $, in particular the best-fit $\dv$ from the void size distribution (solid, red).} 
\label{fig:voidbias_simPBS_compare}
\end{figure*}

In the case of halos, $b_{\rm c} $ can be well described by a constant value on large scales, known as linear halo bias. Here we follow this approach and fit a constant to the low-$k$ plateau of $b_{\rm c} $, as shown in Fig.~\ref{fig:bc_z0_BoxCompare}. 
For the smaller box size, we only consider the bin of radius 7.5 $\MpcOh $, as the other cases in L250 do not reach a plateau in the shown range of $k$ yet. To isolate the dependence of the large-scale bias of voids on their radius, we display in Fig.~\ref{fig:voidbias_simPBS_compare} the best fit large-scale value of $b_{\rm c} $ as a function of $R$ for $z=1$, 0.5 and 0. 
We compare these measurements with the prediction of Eq.~\ref{eq:b1void} assuming different values of $\dv $. Here again, $\dv =-2.8$ significantly underestimates the data, whereas a better (qualitative) agreement is achieved when the best-fit $\dv  $ derived from the void size distribution is used. This is particularly true for voids with $R>20 \MpcOh $ at $z=0$, while the agreement slightly deteriorates at higher redshifts. 
The results from L250 with higher sampling densities are somewhat below the PBS prediction, which is likely due to the non-negligible contribution of sub-voids in the higher sampling densities. 

We now comment on the other sub-samples that we consider.   When only the top-level voids are selected from the void catalogue, with lowest sampling density (0.02 $(\MpcOh)^{-3}$), we find that $b_{\rm c}$ is somewhat larger at low $k$ by roughly $\sim 0.5 $. However, for the high density cases, the trend with sampling density is opposite to the one shown, so the low-$k$ plateau increases with higher sampling densities. 
Restricting ourselves to voids with central density less than $ 0.2 \bar{ \rho}_{\rm m } $ has little effect on the results, except for an increase in  noise due to smaller sample size.


\subsection{The void density profile and its Fourier transform }

The void radial density profile $\rho_{ \rm v}(r) $ describes the distribution of matter conditioned on having a void  center at $r=0$. It can be shown that it is the same as    the cross-correlation function $\xi_{\rm c}(r) $ between void centers and dark matter particles (see e.g.~P.~144 of \cite{Peebles}). Using the conditional form of the correlation function, we can write
\beq
\label{eq:vp_xic}
\rho_{ \rm v }(r) = \bar{ \rho}_{\rm m} [ 1 + \xi_{\rm c} ( r )].
\eeq
So the relative void density profile compared to the mean density $\bar{ \rho}_{\rm m}$,
\beq
\Delta_{\rm v}(r) \equiv \frac{\rho_{ \rm v }(r)  }{ \bar{ \rho}_{\rm m}  } - 1 ,
\eeq
is simply equal to  $\xi_{\rm c} ( r )$. An accurate fitting formula for $\Delta_{\rm v}(r)$ was proposed in Ref.~\cite{HamausSutteretal2014}:
\beq
\label{eq:HSW_profile}
\Delta_{\rm v}(r) = \delta_{\rm cen} \frac{1 - \left( \frac{r  }{ r_{\rm s} } \right)^\alpha  }{1 + \left( \frac{r  }{ R } \right)^\beta  }, 
\eeq
where both the central density fluctuation $ \delta_{\rm cen} $ and the scale radius $ r_{\rm s }$ (the radius at which $\Delta_{\rm v} $ vanishes) are free parameters. As demonstrated in Ref.~\cite{HamausSutteretal2014}, the remaining two parameters $\alpha$ and $\beta$ can be parametrized in terms of $r_{\rm s}  $ and the void radius $R$. These parametrizations have been calibrated at $z=0$ using a sampling density of 0.02 $ ( \MpcOh )^{-3} $. However, since we are probing a wide range of sampling densities and redshifts, we will allow all the four parameters of Eq.~\ref{eq:HSW_profile} to vary freely when we fit them to our simulation data. One should bear in mind that the empirical formula in Eq.~\ref{eq:HSW_profile} is only accurate out to a few times the void radius $R$ and does not capture the large-scale correlation regime.  Although the equivalence between $\Delta_{\rm v} $ and $\xi_{\rm c} ( r )$ is mathematically exact, we prefer to keep separate notations to highlight that $\Delta_{\rm v} $ measurements are often accurate only for relatively small $r$.

\begin{figure*}[!htb]
\centering
\includegraphics[width=\linewidth]{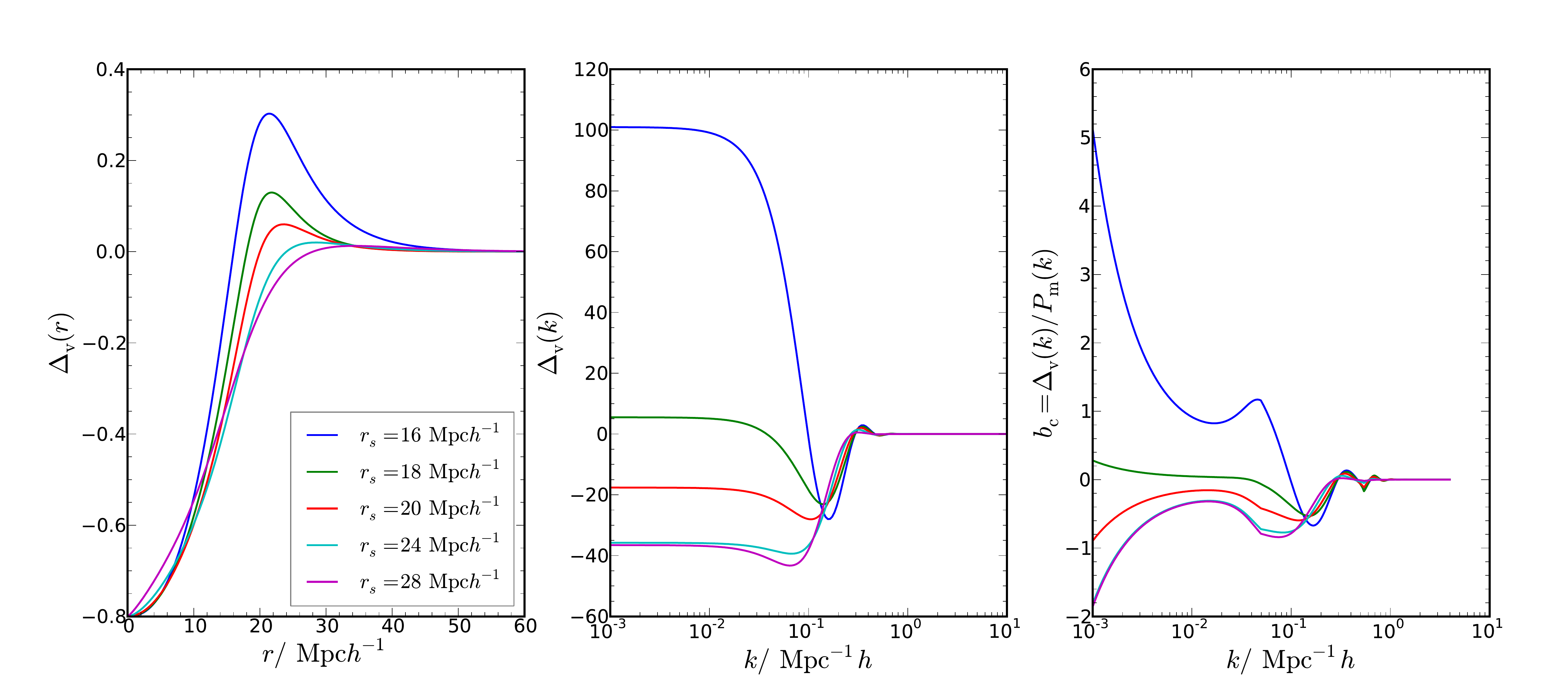}
\caption{ The void density profile in configuration space (left), its Fourier transform (middle), and its Fourier transform divided by the nonlinear matter power spectrum as predicted by Eq.~\ref{eq:HSW_profile}. Here, $R =20  \MpcOh $, $\delta_{\rm cen} = -0.8  $, and $ \alpha $ and $\beta  $ are determined according the parametrization in Ref.~\cite{HamausSutteretal2014}. The values for $r_{\rm s}$ are shown in the inset. }
\label{fig:void_profile_HSW}
\end{figure*}

Before examining the numerical data, we first investigate the properties of the void density profile given in Eq.~\ref{eq:HSW_profile}. For simplicity, we adopt here values for $\alpha $ and $\beta $ derived from the parametrization of Ref.~\cite{HamausSutteretal2014}, keep $\delta_{\rm cen}=-0.8$ fixed and only vary $r_s$. 
In the left-hand panel of Fig.~\ref{fig:void_profile_HSW}, the void density profile from Eq.~\ref{eq:HSW_profile} is shown for five different values of $ r_{\rm s}  $. Note that $r_{\rm s}$ mainly controls the amplitude of the ridge at the void edge (compensation wall). When $r_{\rm s} $ is small (large) compared to $R$, the ridge is high (low). The Fourier transform of $\Delta_{\rm v}(r) $ is shown in the middle panel of Fig.~\ref{fig:void_profile_HSW}. It is formally equal to the void-matter cross-power spectrum
\beq
\label{eq:Pc_FourierTransform}
P_{\rm c} (k ) = \int  \frac{  4 \pi r^2 dr  } {  ( 2 \pi )^3  }  \frac{ \sin ( k r) }{ kr } \Delta_{\rm v}(r) .
\eeq
However, as the empirical profile in Eq.~\ref{eq:HSW_profile} is expected to be accurate only up to a few times of $ R $, the low-$k$ regime in its Fourier transform should not be trusted. Although the amplitude of $\Delta_{\rm v}  $ becomes quite small outside a few void radii, it may still yield sizable contributions to the Fourier transform integral in Eq.~\ref{eq:Pc_FourierTransform}. 
To facilitate the comparison with our simulation results, we normalize $\Delta_{\rm v}( k) $ with respect to the nonlinear dark matter power spectrum in the right-hand panel of Fig.~\ref{fig:void_profile_HSW}. This furnishes an estimate for the void-matter cross-bias $b_{\rm c}(k) $. The resulting predictions for $b_c$ are qualitatively similar to those displayed in Fig.~\ref{fig:bc_z0_BoxCompare}, except for a scale dependence at low $k$ which is particularly large for voids with high ridges.
This strong scale dependence as $k\to 0$ originates from the fact that Eq.~\ref{eq:HSW_profile} does not properly take into account large-scale correlations. 





We also measure the spherically averaged void density profile in our simulations by stacking voids in different radius bins in Fig. 7 (for details, see \cite{Sutteretal2014}). 
We find  good agreement between L1500 and L250 for the identical sampling density of 0.02 $(\MpcOh)^{-3} $. Voids from higher sampling densities exhibit lower ridges at fixed void radius, in agreement with Ref.~\cite{Sutteretal2013}. Overall, Eq.~\ref{eq:HSW_profile} offers a very good description of the simulation data. 

\begin{figure*}[!htb]
\centering
\includegraphics[width=\linewidth]{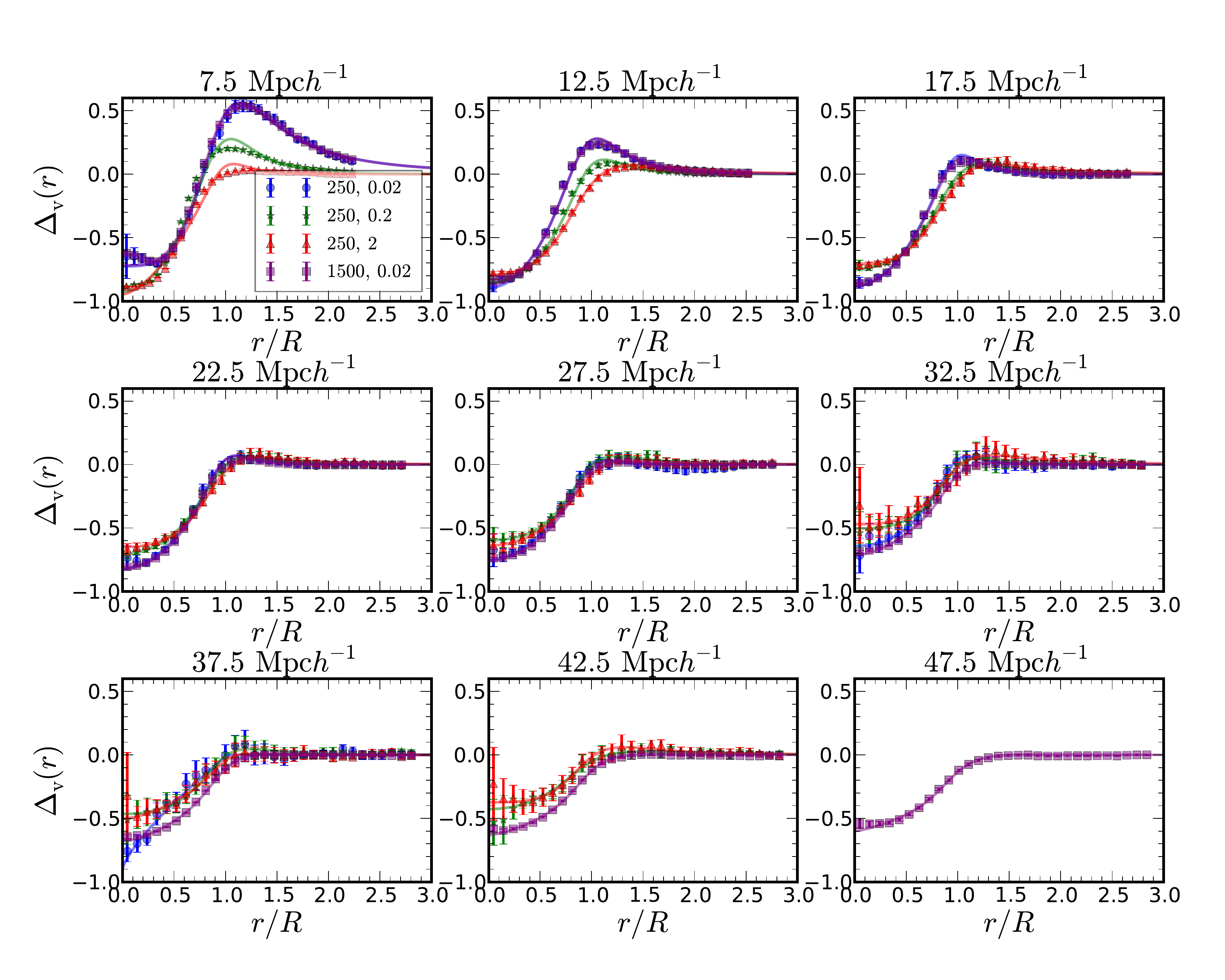}
\caption{ Void density profiles for different void radii at $z=0$ from L1500 with sampling density  0.02 $ (\MpcOh)^{-3}   $ (square, violet) and L250 with sampling densities 0.02 (circle, blue), 0.2  (star, green), and 2  $(\MpcOh)^{-3}$ (triangle, red). Solid lines show the best fits using Eq.~\ref{eq:HSW_profile} (same colors). }  
\label{fig:void_profile_4parafit_z0}
\end{figure*}



\begin{figure*}[!htb]
\centering
\includegraphics[width=\linewidth]{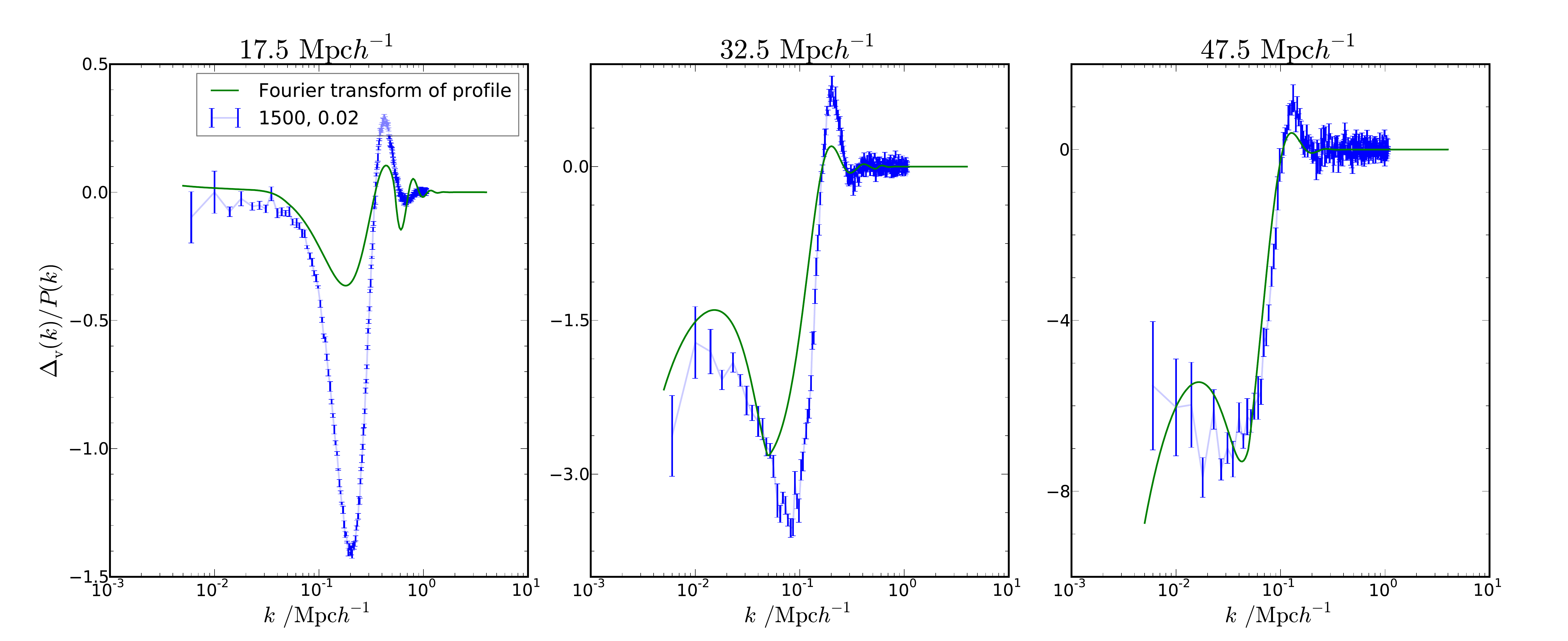}
\caption{ Comparison of the void-matter cross bias $b_{\rm c} $ from the L1500 simulation (blue with  error bars) with the Fourier transform of the corresponding best-fit void density profile divided by the matter power spectrum (green lines) at $z=0$.}
\label{fig:void_profile_4para_z0_k_Pknorm_sample}
\end{figure*}

Because the void density profile is measured accurately out to a few void radii only, and small inaccuracies can end up as large deviation in Fourier space, it is not clear how well one can predict the cross power spectrum using the void profile fitted by Eq.~\ref{eq:HSW_profile}. In practice, we extrapolate a small-scale quantity to large scales in order to predict the power spectrum down to low values of $k$. In Fig.~\ref{fig:void_profile_4para_z0_k_Pknorm_sample} we compare the $z=0$ void cross bias $b_{\rm c} $ from the  L1500 simulation with that obtained from a Fourier transform of the best-fit void density profile shown in Fig.~\ref{fig:void_profile_4parafit_z0}. 
As expected, the agreement is only qualitative, although it improves as the void size increases. 
At low $k$, the profile from Eq.~\ref{eq:HSW_profile} often causes a strong scale dependence.  
Furthermore, the amplitude of the first two prominent oscillations beyond the low-$k$ plateau is underestimated, even though their shape is reproduced. 
The agreement improves for larger voids, essentially because their density profile  is  measured out to larger distances $r$, or equivalently, lower values of $k$. 


Before the end of this section, we would like to discuss an interesting possibility that the cross bias parameter may exhibit scale dependence at very low $k $.  At large $r$, the void-matter cross-correlation function is expected to be proportional to the dark matter correlation function. Therefore, we can split Eq.~\ref{eq:Pc_FourierTransform} into two contributions, 
\beqa 
\label{eq:Pc_split}
P_{\rm c} (k )& = &   \int_0^{r_*}  \frac{ 4 \pi r^2 dr }{( 2\pi )^3  } \frac{ \sin kr}{ kr} [  \Delta_{\rm v} ( r ) - b_1 \xi_{\rm m}(r) ]  \nn   \\
&+&  b_1  \int_0^\infty   \frac{ 4 \pi r^2 dr }{( 2\pi )^3  } \frac{ \sin kr}{ kr}  \xi_{\rm m}(r), 
\eeqa
where the scale $r_*$ is determined with the simplifying assumption that $ \Delta_{\rm v}(r) = b_1 \xi_{\rm m}(r) $ for $r > r_* $.  The magnitude of  $r_*$  may be taken  to be    a few void radii.  Hence, $ \Delta_{\rm v}(r) - b_1 \xi_{\rm m}(r) $ does not vanish if void bias is different from a simple $k$-independent linear contribution.
Since at small $k$ we have $\mathrm{ sin  }(kr)/kr \sim 1 $, the first integral on the right-hand side of Eq.~\ref{eq:Pc_split} yields a constant in the limit $k\to 0$:
\beq
P_{\rm c} (k) = \mathrm{const.} + b_1 P_{\rm m }(k), 
\eeq
or equivalently
\beq
\label{eq:bc_scale_dependent}
b_{\rm c}(k) = \frac{ \mathrm{const.} }{ P_{ \rm m}(k) }  + b_1. 
\eeq
The first term in Eq.~\ref{eq:bc_scale_dependent} can generate a residual $k$-dependent bias at very low $k$. 
Our argument is similar to that given in Ref.~\cite{ScherrerWeinberg98}, where the existence of white noise power in the low-$k$ {\it auto-power spectrum}  is expected if biasing is nonlinear.  Fig. \ref{fig:bc_z0_BoxCompare} does not show  any clear  residual scale dependence at low $k$, which would indicate the constant in Eq.~\ref{eq:Pc_split} to be non-zero. We can thus safely ignore it in this paper. A more decisive check can be made if the measurement is extended to lower $k$, such as  $ k \lesssim 10^{-3} \hOMpc $.

\subsection{The void-matter cross-power spectrum and its inverse Fourier transform.}
In this section, we follow a different approach and first attempt to describe the void-matter cross-power spectrum in Fourier space with a physically motivated formula. We subsequently use it to predict the void density profile in configuration space. 
In order to write down an expression for $b_{\rm c}(k) $, we take advantage of the close connection between voids and minima of the linear density field, as already discussed in SvdW. Since the whole peak formalism straightforwardly applies to the description of initial minima, we expect the void linear cross-bias to exhibit a $k^2$ dependence \cite{bbks}. Furthermore, since voids trace a smoothed version of the mass density field, while $b_{\rm c}$ is defined relative to the unsmoothed mass distribution, we must also include a filter function. For these reasons, we consider the following parametrization:
\beq
\label{eq:k4Gauss_bc}
b_{\rm c} (k ) = ( b_0 + b_2 k^2 +  b_4 k^4 ) \exp \Big[ - \frac{1}{2} ( k R_{\rm G} )^2 \Big], 
\eeq
where $b_0 $,  $b_2$,  $b_4$ and $R_{\rm G} $ are free parameters. Here, $(b_0 + b_2 k^2)$ times the Gaussian is the Fourier transform of the peak profile derived in \cite{bbks}, which naturally arises from the peak constraint. Note that exactly the  same linear bias appears in the peak {\it auto-power spectrum} \cite{vd08}. 
We have added a $k^4$ piece in order to better reproduce the large oscillatory features. The excursion set peak (ESP) prescription of \cite{paranjapesheth,PSD13}, which generates additional $k$-dependence through the first-crossing constraint \cite{vdgongriotto,biagettietal}, would provide a more consistent description of $b_{\rm c}(k)$ . However, an extension of the ESP formalism to voids is beyond the scope of this paper. We will thus stick to the simple formula in Eq.~\ref{eq:k4Gauss_bc}. 

\begin{figure*}[!htb]
\centering
\includegraphics[width=\linewidth]{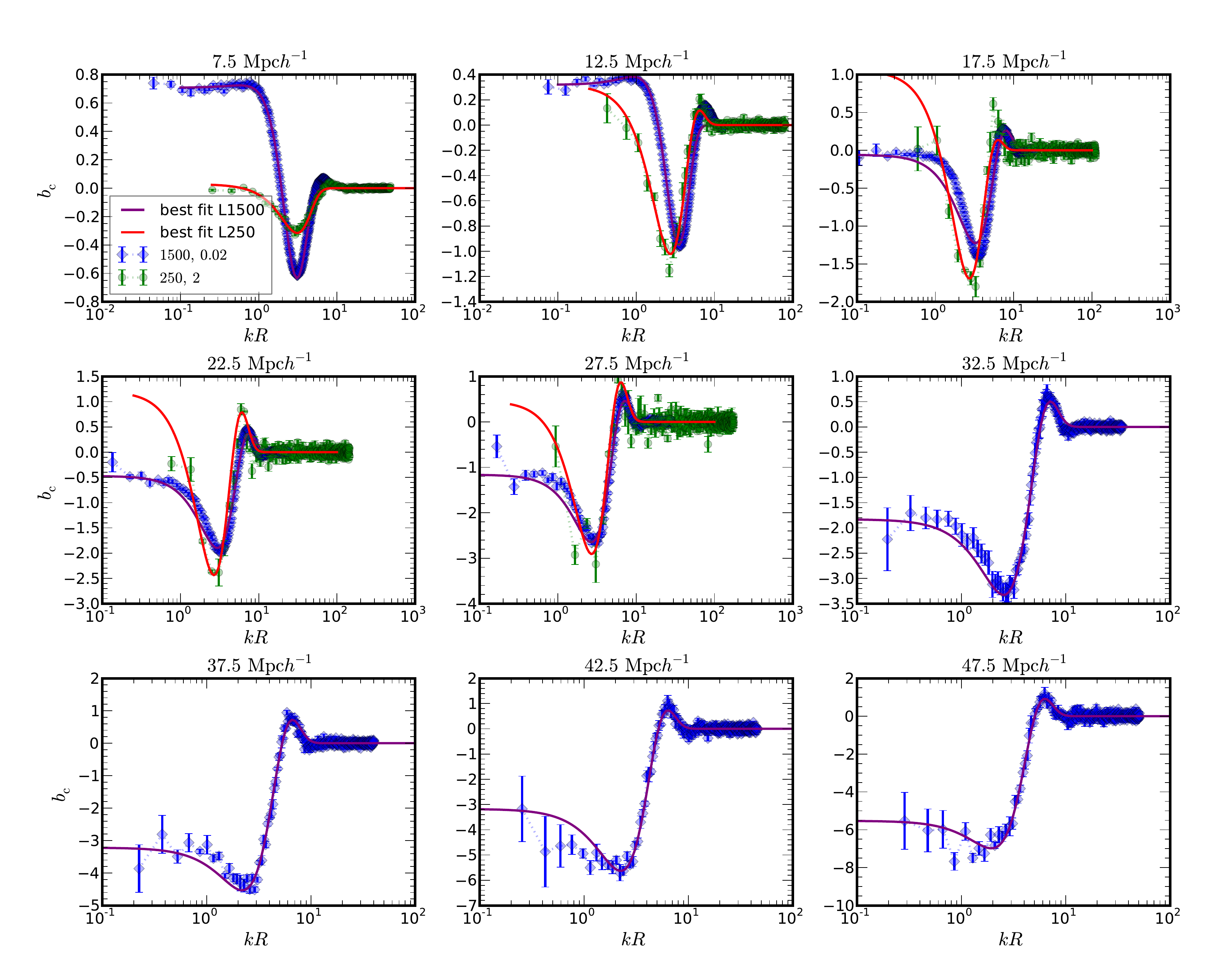}
\caption{ Void-matter cross bias $b_c$ for different void sizes from L1500 (diamond, blue) and L250 with sampling density 2 $(\MpcOh)^{-3} $ (circle, green) and its best fits using Eq.~\ref{eq:k4Gauss_bc} (violet and red lines, respectively) at $z=0$. }
\label{fig:bck_k4GaussFit_z0}
\end{figure*}

In Fig.~\ref{fig:bck_k4GaussFit_z0} we show $b_{\rm c}(k)$ at $z=0$ from L1500 and L250 with sampling density 2 $(\MpcOh)^{-3}$ together with the best fit obtained from Eq.~\ref{eq:k4Gauss_bc}. We are able to fit the simulation data for voids of various sizes and sampling densities very well, with the caveat that we first determine $b_0$ separately by fitting a constant at low $k$ before fitting the entire data to constrain the other parameters. 
We indeed found that, if we fit Eq.~\ref{eq:k4Gauss_bc} to $b_{\rm c} $ directly, the best fit often overshoots at low $k$ and lies above the low-$k$ plateau. 
Our two-step fitting procedure alleviates this problem.
Note that a second trough of smaller amplitude beyond the maximum of $b_{\rm c}(k)$ is visible in some cases in Fig.~\ref{fig:bck_k4GaussFit_z0}, but to fit this additional feature higher order terms, such as $k^6$, have to be included in Eq.~\ref{eq:k4Gauss_bc}.


In the spirit of the previous Section, we can try to predict the void density profile using information from Fourier space. We can carry out the inverse Fourier transform of $P_{\rm c}=b_{\rm c }P_{\rm m} $ using the best fit from Fig.~\ref{fig:bck_k4GaussFit_z0}. However, we find that generally the void density profile in the void interior is poorly reproduced, with the density contrast much lower than the measured one. In particular, the constraint that $ \delta \ge -1 $ is often not satisfied. This suggests that the high-$k$ structure of $b_{\rm c}$ is important to describe the interior of the void density profile.  As expected, the discrepancy is largest close to the void center and diminishes towards larger distances, especially for bigger voids for which the ``peak'' approximation should be most accurate.

\section{Void power spectrum and exclusion }
\label{sec:VoidPk_exclusion}

The void auto-power spectrum is more closely related to observational data than the void-matter cross-power spectrum, because voids can also be defined in the spatial distribution of galaxies without knowledge of the underlying  dark matter density field. As these voids are biased tracers of the mass, this allows one to infer information about the dark matter power spectrum. Furthermore, since  the void size distribution and the void bias are sensitive to the definition of voids,  it is useful to  regard them as nuisance parameters in the void auto-power spectrum that can be marginalized over, similarly to what is being done in galaxy clustering analyses. In this section we investigate this approach, but also compare our numerical results to the PBS predictions.

In analogy to halos, the void auto-power spectrum is affected by shot noise due to the discrete nature of voids. Poisson shot noise is straightforward to model, its contribution to the void power spectrum is scale independent and is given by
\beq
\label{eq:PPoi}
P_{\rm Poi} = \frac{ 1 }{ (2 \pi)^3 \bar{n}_{\rm v} }. 
\eeq

Voids are also biased tracers of the density field. On large scales, it is usually sufficient to consider only linear bias. However, as we will extend our analysis to $k \gtrsim 0.1 \hOMpc$, we have to consider higher orders as well. In fact, as we shall see below, even at low $k$ linear bias may not be sufficient. For simplicity and comparison purposes, we begin with a linear bias model, which we will further extend using the renormalized bias approach \cite{McDonaldReBias} to include higher order bias  up  to the 1-loop order.

\begin{figure*}[!htb]
\centering
\includegraphics[width=\linewidth]{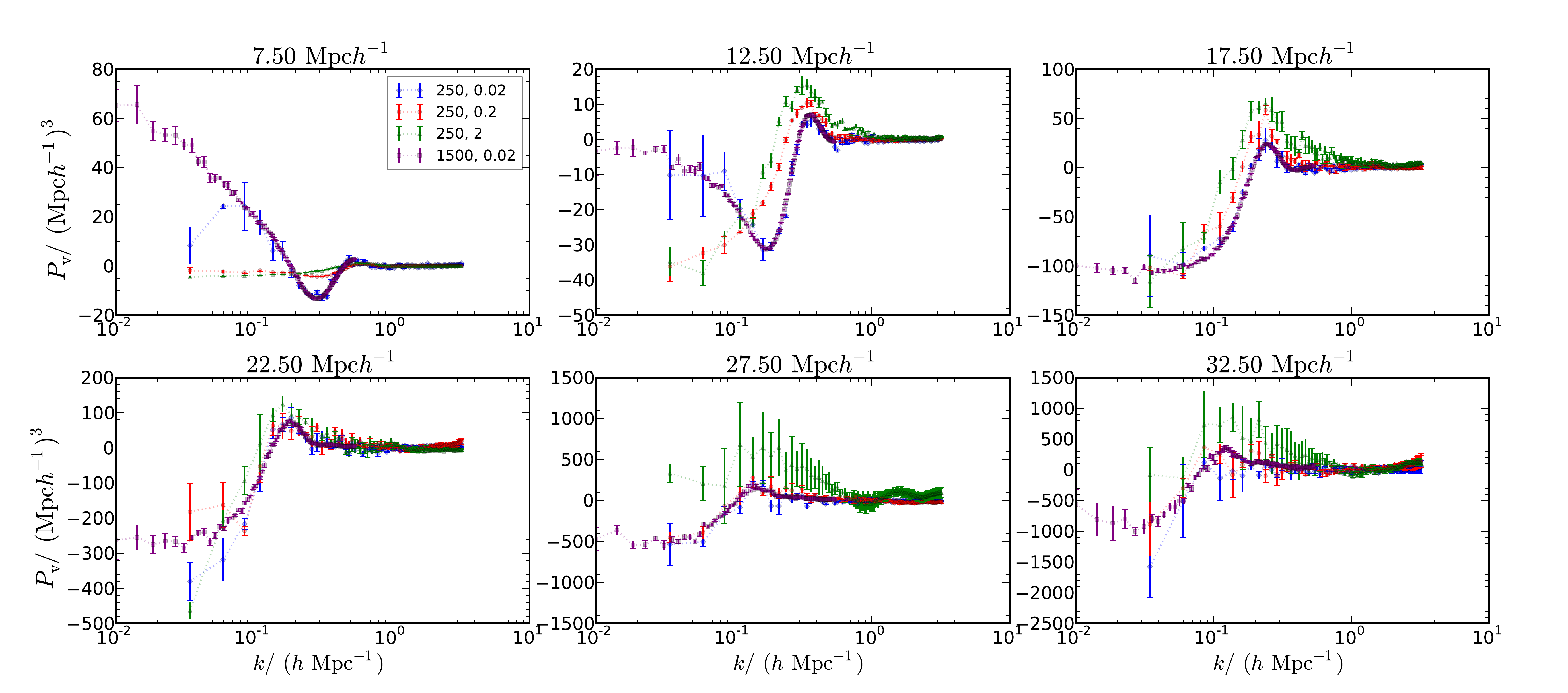}
\caption{  Void auto-power spectrum from L1500 (square, violet) and L250 with sampling densities 0.02 (diamond, blue), 0.2 (circle, red), and 2 $ (\MpcOh)^{-3}  $ (triangle, green) for voids of different size at $z=0$. Poisson shot noise has been subtracted. }
\label{fig:Pkv_sampling_density_check}
\end{figure*}

The sampling-density dependence of the void auto-power spectrum at $z=0$ is shown in Fig.~\ref{fig:Pkv_sampling_density_check} for different bins in void size upon subtracting the Poisson shot noise. The results obtained from L1500 and L250 with the same sampling density of 0.02 $( \MpcOh)^{-3}$ are consistent with each other. However, the power spectra of small voids from different sampling densities exhibit larger discrepancies, which are again likely caused by their high sub-void fraction. 
As the void size increases, differences among various sampling densities are reduced. 
The power spectrum of small voids ($R \lesssim 15 \MpcOh $) features a bump at low $k$ for the lowest sampling density ($0.02 (\MpcOh)^{-3}$). 
This feature is more significant and persists for larger void radii ($R \lesssim25 \MpcOh$) when we restrict the sample to top-level voids. Furthermore, for large voids and higher sampling densities, the power spectrum reaches a negative plateau at low wavenumber. 
These findings are weakly sensitive to the exact value of the central density cut. 

\subsection{ Void exclusion }

Because voids are generally much more extended than halos, void exclusion plays an important role in modeling the void power spectrum. To this end we adopt the hard-sphere approximation commonly used in statistical mechanics to describe simple liquids and non-ideal gases (see e.g.~\cite{HansenMcDonald,Torquato}). Recently, hard-sphere models have also been applied to model halo exclusion \cite{Smithetal2007,Baldaufetal2013,ChanScoccimarro2014}. A system of hard spheres exhibits correlations that are purely induced by their finite size. In an  ensemble  of identical hard spheres of diameter $D$ and number density $\bar{n}$ in equilibrium, the correlation function is accurately described by the so-called Percus-Yervick equation.  The corresponding power spectrum is given by \cite{HansenMcDonald,Torquato,ChanScoccimarro2014}
\beq
\label{eq:OrnsteinZernike_k}
P_{\rm HS}( k ) =   \frac{  c( k ) }{ 1  -  (2 \pi)^3  \bar{n}  c( k ) }.
\eeq
with  
\beqa
\label{eq:c_k}
c( k ) &=& - \frac{  D^3  }{2 \pi^2 q^3 } \Big\{ a_1 ( \sin q - q \cos q )
+ \frac{ 6 \eta a_2  }{q } \big[  2 q \sin q  \nn \\
&+&  ( 2 - q^2 ) \cos q - 2  \big]
+ \frac{ \eta a_1 }{ 2 q^3  }\big[  4q ( q^2 -6) \sin q   \nn \\ 
& -& ( 24 - 12 q^2 + q^4 ) \cos q + 24    \big]
\Big\} , 
\eeqa
where  $ q= kD $, $\eta $ is the packing fraction,
\beq
\eta =\frac{ \pi \bar{n} D^3  }{ 6 },
\eeq
and $a_1$ and $a_2$ are given by
\beq
a_1 = \frac{(1+ 2 \eta )^2  }{ (1 -\eta )^4 }, \quad a_2 = - \frac{ (1 + \eta /2 )^2 }{(1- \eta )^4  }. 
\eeq

When $\eta\ll1$, the hard-sphere correlation is well approximated by a top-hat window in configuration space \cite{HansenMcDonald,Torquato,Smithetal2007}
\begin{equation}
\label{eq:h_naive}
 \xi_{\rm TH}(r)  =  \left\{
 \begin{array}{cl}
 -1          &       \text{for }         r<D , \\
   0         &  \text{for }    r \geq D .  \\
 \end{array} \right.
 \end{equation}
Its Fourier transform is 
\beq
\label{eq:NaiveHardSphere}
P_{\rm TH}(k) = \frac{ D^3 }{ 2 \pi^2 q^3  } ( q \cos q - \sin q  ).
\eeq

Due to the sharp transition between the exclusion zone and the outer region, hard-sphere models feature strong ringing in Fourier space. By contrast, voids are not perfect spheres. Also binning of voids in radius also reduce the sharpness of the effective radius.  Because of these, a soft-sphere model that smoothly interpolates the transition region in the correlation function from $-1$ to $0$ may be  more realistic. For example, we can use the function $\tanh(r) $ for that purpose, which is well known in studies of domain walls. Thus, in configuration space we write
\beq
\label{eq:xi_tanh}
\xi_{\tanh}( r ) =   \frac{ 1 }{ 2 }  \Big[  \tanh \Big( \frac{ r - D }{\sigma_{\rm t} }   \Big)  - 1   \Big]  ,  
\eeq
where $\sigma_{\rm t}$ is the width of the transition region. Unfortunately, no simple analytic form for the Fourier transform of this function exists. 
Multiplying $P_{\rm HS} $ or  $P_{\rm TH} $ by a Gaussian damping factor of width $\sigma_{\rm G}$  in Fourier space  achieves a similar smoothing,
\beq
W_{\rm G}(k) = \exp \Big[ -  \frac{  1}{ 2 } (  \sigma_{\rm G} k )^2   \Big].
\eeq

In the case of halos, the residual power spectrum $P_{\rm res} $ is often considered in order to study the large-scale noise contribution to the halo power spectrum $P_{\rm h} $ (e.g.~\cite{SeljekHamausDesjacques2009,HamausSeljeketal2010,ChanScoccimarro2014}):
\beq
\label{eq:Pres_standard}
P_{\rm res}(k) = P_{\rm h}(k) - b_1^2 P_{\rm m}(k), 
\eeq
where $b_1$ is the cross-bias parameter in the low-$k$ limit.  
In fact, this can be regarded as a definition of ``noise'' in the auto-power spectrum.  As the measured $b_{\rm c}  $ only qualitatively agrees with the PBS predictions,  we will not follow Eq.~\ref{eq:Pres_standard}, but use $b_1$  as a free parameter below.


\subsection{ Linear bias model} 

We begin with the simple linear bias model, in which
\beqa
\label{eq:Pv_linear}
P_{\rm v} &=& P_{1,1} + P_{\rm excl},    \\ 
P_{1,1} &=&  b_1^2 P_{\rm m} ,
\eeqa
where $ P_{\rm m} $ denotes the nonlinear dark matter power spectrum. We shall consider five different approximations to $P_{\rm excl}$ as listed in Table \ref{tab:Pexcl_model}. 
Models 1 and 2 contain two free parameters, $b_1$ and $D$, while the other ones include one additional smoothing parameter ($\sigma_{\rm t}  $ or $\sigma_{\rm  G} $).    Also note that $P_{\tanh} $ is computed via numerical Fourier transform of Eq.~\ref{eq:xi_tanh}. 

\begin{table}[!htb]
\caption{ Different models for $P_{\rm excl} $. }
\label{tab:Pexcl_model}
\begin{ruledtabular}
\begin{tabular}{c|l}
Model \# & $P_{\rm excl}$  \hspace{175pt}             \\
\hline 
1        & $P_{\rm HS} (k ; D)$                       \\
2        & $P_{\rm TH}(k; D)$                         \\
3        & $P_{\tanh} (k; D, \sigma_{\rm t})$         \\  
4        & $W(k; \sigma_{\rm G} ) P_{\rm HS} (k ; D)$ \\
5        & $W(k; \sigma_{\rm G} ) P_{\rm TH} (k ; D)$ \\
\end{tabular}
\end{ruledtabular}
\end{table}

\begin{figure*}[!htb]
\centering
\includegraphics[width=\linewidth]{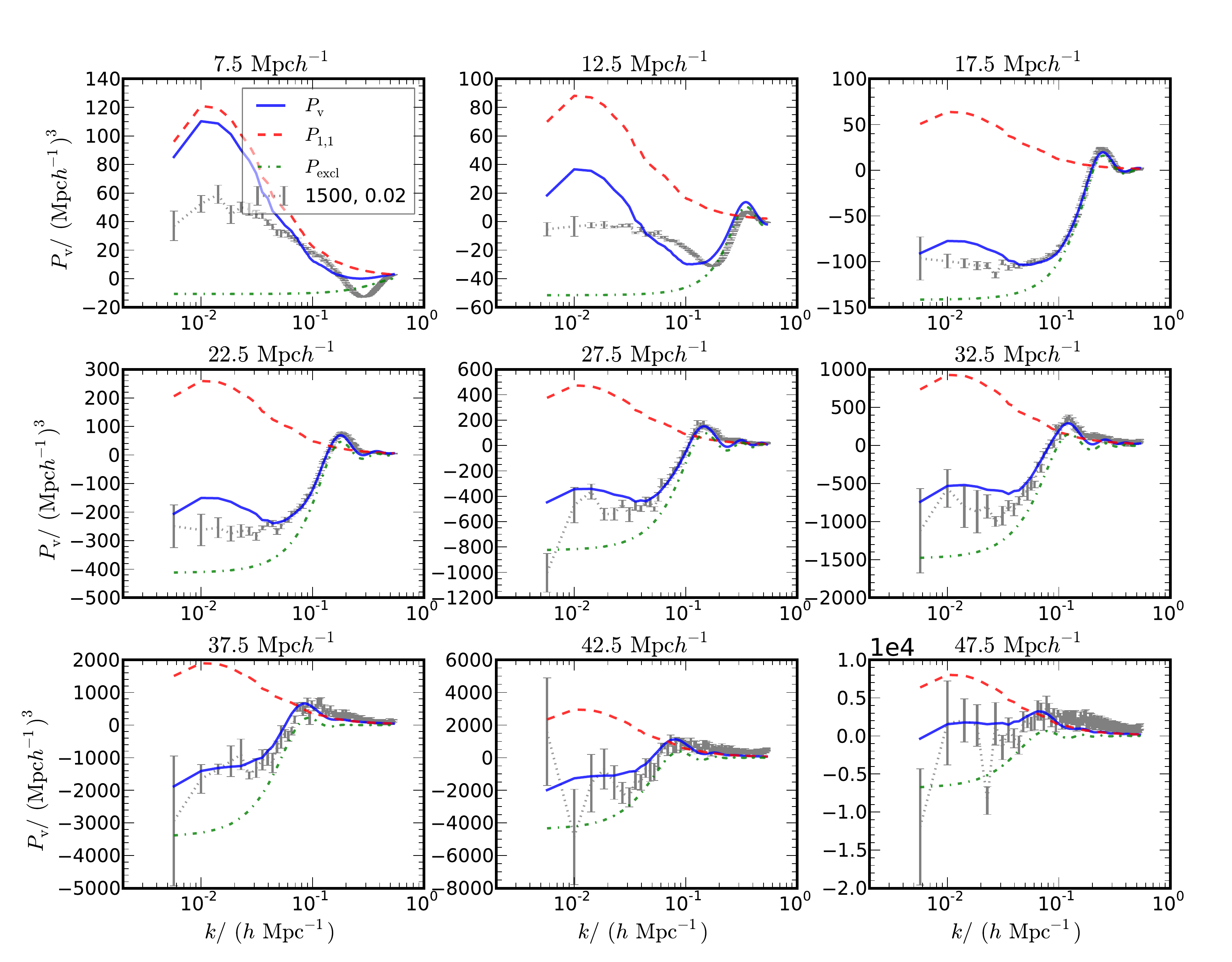}
\caption{ Best fits from Eq.~\ref{eq:Pv_linear} (solid, blue) to the void auto-power spectrum from L1500 at $z=0 $ (grey data points with error bars). The individual components of the best fit are also shown: the linear bias term $P_{1,1}  $ (dashed, red) and the void-exclusion term $P_{\rm excl} $ (dotted-dashed, green). Here, model 1 is used for  $P_{\rm excl}   $ and Poisson shot noise has been subtracted from the data. } 
\label{fig:Pv_b1_HS1_components}
\end{figure*}

In Fig.~\ref{fig:Pv_b1_HS1_components} we show the best fits of Eq.~\ref{eq:Pv_linear} to the void auto-power spectrum from L1500 for different void radius bins at $z=0$. We also plot the individual terms from Eq.~\ref{eq:Pv_linear} separately to highlight their relative importance. Poisson shot noise has been subtracted from the numerical results, and only data points up to $k = 0.2 \hOMpc$ are included in the fit. For clarity, only results obtained with model 1 are shown, as the others all lead to similar results. 
Eq.~\ref{eq:Pv_linear} gives a poor fit for the smallest voids (notably $R=7.5 $ and $12.5 \MpcOh  $), but describes the power spectra for larger voids reasonably well. The exclusion term plays a central role, especially in the low-$k$ regime, where $P_{1,1}$ and $P_{\rm excl}$ are of opposite sign, but the magnitude of the exclusion term is comparable or even larger than that of the $b_1$-term. 
Towards smaller scales the fits are not performing as well, firstly because linear bias is not sufficient on those scales, and secondly because the oscillations from the hard-sphere model are too strong. Although the high-$k$ wiggles in model 3, 4 and 5 are damped, their best fits do not have the correct amplitude.

\begin{figure*}[!htb]
\centering
\includegraphics[width=\linewidth]{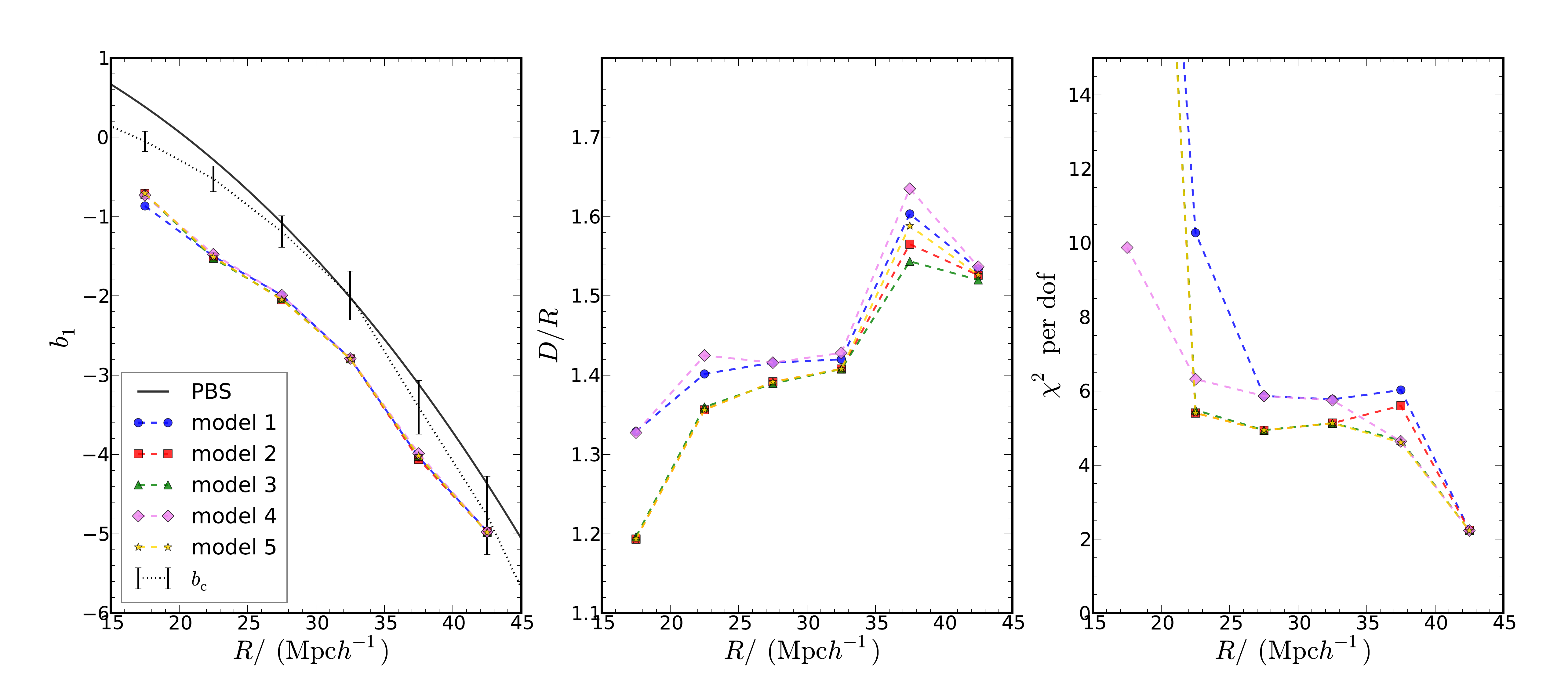}
\caption{ The best-fit values for $b_1$ and $D$ using Eq.~\ref{eq:Pv_linear} to fit the void auto-power spectrum and the corresponding $\chi^2  $ per degree of freedom. The results from model 1 (circle, blue), 2 (square, red), 3 (triangle, green), 4 (diamond, violet) and 5 (star, yellow) are shown. The solid line depicts $b_1 $ as predicted from the PBS formalism. We also show the large-scale measurement of $b_{\rm c} $ (dotted, black) for comparison.   }
\label{fig:bestfit_b1}
\end{figure*}

The best-fit values for $b_1 $, $D$ (in unit of void radius $R$) and the $ \chi^2 $ per degree of freedom are plotted in Fig.~\ref{fig:bestfit_b1} as a function of $R$ for our five models, they all yield fairly similar results. For the sake of comparison, we also plot  $b_1  $ from the PBS formalism, i.e. Eq.~\ref{eq:b1void} with the best fit $\dv  $ from the void size distribution.  Its $R$ dependence qualitatively agrees with the data, but its amplitude is systematically larger by unity or so.  
On the other hand, the large-scale $b_{\rm c} $ measurements agree with the PBS results much better. We will comment on this issue more in next section.  The $\chi^2  $ per degree of freedom, with values of $\sim 5$ for the intermediate range of void sizes, reflects the rather poor fits obtained with the linear bias approximation. It is even worse at small $R$, as apparent from Fig.~\ref{fig:Pv_b1_HS1_components}.

Looking at the best-fit values for $D$, we find that models 1 and 4 on the one hand, and models 2, 3 and 5 on the other hand yield very similar results. This is not surprising, as model 4 differs from model 1 only by a Gaussian damping factor, which only affects relatively high $k$, while models 2, 3 and 5 are all variants of the top-hat window. Overall, all models share a similar trend. If $D$ is taken to be the Eulerian size of a void, then $D/R $ should be equal to 2. However, if we use the Lagrangian size for voids, according to Eq.~\ref{eq:RL_R_relation} we expect $D/R $ to be about 1.16. In Fig.~\ref{fig:Pv_b1_HS1_components}, $D$ is closer to the Lagrangian estimate at low $R$, and gets closer to the Eulerian estimate for larger void radii.


The model fails for the small voids with  $R\ \lesssim 15 \MpcOh $ whose power spectrum, unlike voids of larger radii, increases as $k$ decreases and eventually becomes positive. Fig.~\ref{fig:Pv_b1_HS1_components} also suggests that the hard-sphere  exclusion model does not work for this  kind small voids. In fact, even the inclusion of nonlinear bias, which will be discussed shortly, does not improve the agreement noticeably. However, because the  model of Eq.~\ref{eq:Pv_linear} involves biasing and exclusion, it is unclear which one is the culprit for the failure. It would be easier to disentangle them in the configuration space correlation function, where exclusion effects are confined at short distances while the (linear) biasing is important at large $r$.


\subsection{ Renormalized bias model} 

Fig.~\ref{fig:Pv_b1_HS1_components} clearly shows that linear bias is not sufficient to model the data accurately and one must consider higher orders. While in this work we will restrict ourselves to the local bias model \cite{FryGaztanaga}, we will include contributions up to third order,  
\beq
\delta_{\rm v} = b_1 \delta + \frac{ b_2}{2 } \delta^2 +  \frac{b_3}{ 6 } \delta^3, 
\eeq
where $b_2 $ and $b_3$ are the second and third order bias parameters~\footnote{ $b_2$ here should not be confused with the coefficient of $k^2$ in Eq.~\ref{eq:k4Gauss_bc}.  }.
We can now make use of the renormalization procedure of Refs.~\cite{McDonaldReBias,JeongKomatsu09,sjd}. In plain words, the standard 1-loop expansion of the halo power spectrum with local bias yields four terms, see e.g. Eq.(12)--(15) and the corresponding diagrammatic expansion in \cite{ChanScoccimarro2012}. However, two of these terms are simply proportional to $P_{\rm m} (k)$, so by absorbing them in $P_{1,1} $ they do not appear explicitly anymore. In the language of \cite{McDonaldReBias}, these terms renormalize $b_1$. As $b_3$ only appears in  one of these two terms,  it  does not explicitly appear at 1-loop order. Hence, there are effectively only two new terms
\beqa
P_{2,11} & =& 2 b_1 b_2 \int d^3 q F_2( \mb{q}, \mb{k} - \mb{q} ) P_{\rm L} (q) P_{\rm L} (|\mb{k} - \mb{q} |),  \\
P_{11,11} &=& \frac{ b_2^2   }{ 2 } \int d^3 q P_{\rm L} (q)  P_{\rm L} (|\mb{k} - \mb{q} |),
\eeqa
and one additional parameter, $b_2$.
Here, $ P_{\rm L} $ is the linear matter power spectrum and $F_2$ denotes the kernel
\beq
F_2 ( \mb{q}, \mb{p} ) = \frac{5}{7} + \frac{1}{2} \mu \Big( \frac{q}{p}+  \frac{p}{q}  \Big) + \frac{2}{7} \mu^2,  
\eeq 
where $\mu = \hat{\mb{q} } \cdot  \hat{\mb{p} }  $. 

\begin{figure*}[!htb]
\centering
\includegraphics[width=\linewidth]{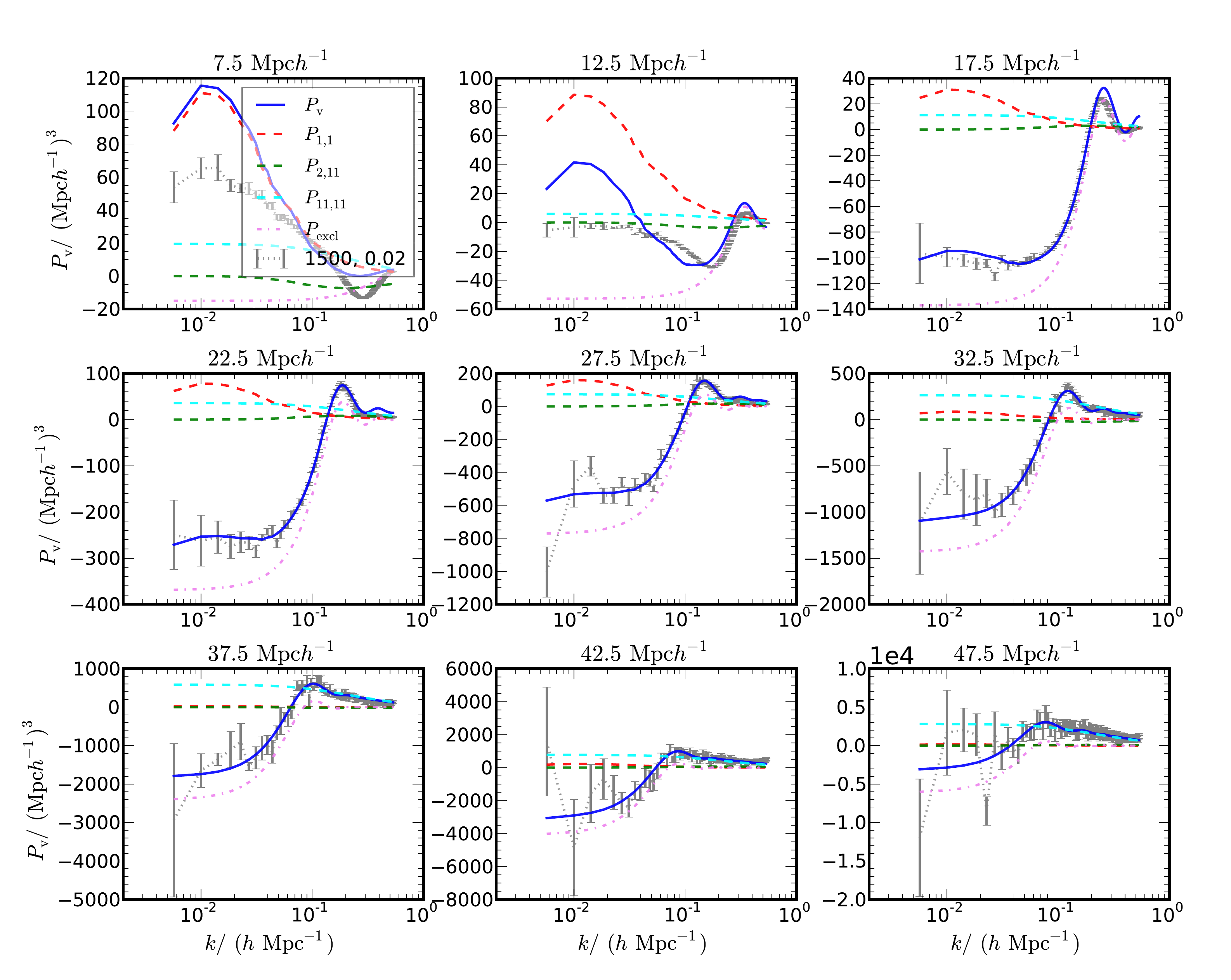}
\caption{  Best fits from Eq.~\ref{eq:Pv_b1b2} (solid, blue) to the void auto-power spectrum from L1500 at $z=0 $ (gray data points with error bars). The individual components of the best fit are also shown: the linear bias term $P_{1,1}  $ (dashed, red), $ P_{2,11} $ (dashed, green), $ P_{11,11} $ (dashed, cyan), and the void-exclusion term $P_{\rm excl} $ (dotted-dashed, violet). Here, model 4 is used for  $P_{\rm excl}   $ and Poisson shot noise has been subtracted from the data.  }
\label{fig:Pv_b1b2_HS1_components}
\end{figure*}

Therefore, we shall adopt the following model for the void power spectrum:
\beq
\label{eq:Pv_b1b2}
P_{\rm v} = P_{1,1} + P_{2,11} + P_{11,11}   + P_{\rm excl}
\eeq
Fig.~\ref{fig:Pv_b1b2_HS1_components} displays the resulting best fits obtained after the inclusion of the $b_2$ terms. Including the quadratic bias significantly improves the agreement with the numerical data, except for the smallest voids ($R=7.5$ and 12.5 $\MpcOh$), for which the fits are still poor.  In this plot, model 4 is used to account for exclusion effects, although the results do not depend on the exact form of $P_\text{excl}$ when $R\gtrsim 25\MpcOh$.

Closer inspection of the individual components of the best-fit power spectrum reveals that the term $P_{2,11} $ is negligible for the entire range of $k$ shown. On the other hand, we find that $P_{11,11} $ (which is nearly constant at low $k$ and thus has been coined ``shot noise renormalization term''  in \cite{McDonaldReBias}) is comparable to, or even larger than $ P_{1,1}$ for the biggest voids. This term drives the improvement of the fit to the void power spectrum compared to the previous linear model. 

\begin{figure*}[!htb]
\centering
\includegraphics[width=\linewidth]{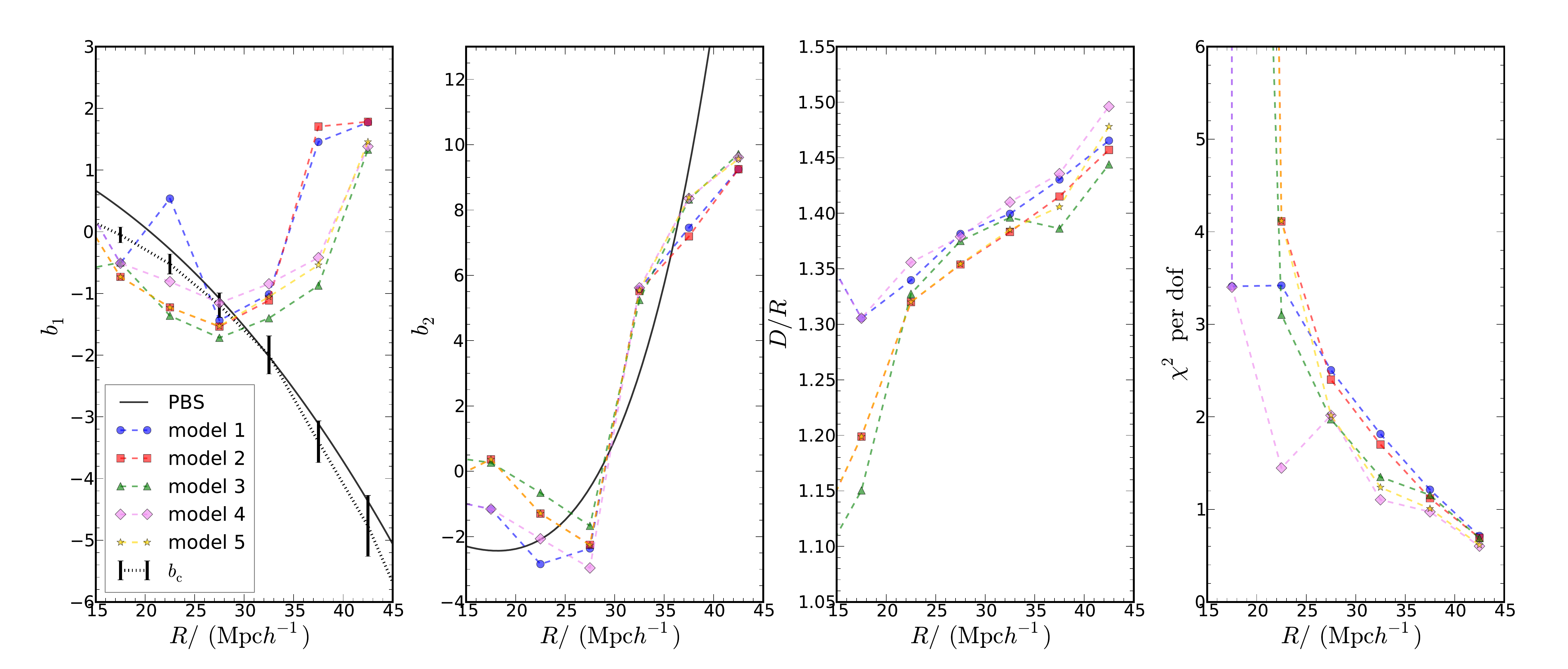}
\caption{ The best-fit values for $b_1$, $b_2$ and $D$ using Eq.~\ref{eq:Pv_b1b2} to fit the void auto-power spectrum and the corresponding $\chi^2  $ per degree of freedom. The results from model 1 (circle, blue), 2 (square, red), 3 (triangle, green), 4 (diamond, violet) and 5 (star, yellow) are shown. The solid lines depict $b_1 $ and $b_2 $ as predicted from the PBS formalism.  The large-scale measurement of $b_{\rm c}$ is also plotted (dotted, black).    }
\label{fig:bestfit_b1b2}
\end{figure*}

The best-fit parameters $b_1$, $b_2$, $D$, and the $\chi^2  $ per degree of freedom  are shown in Fig.~\ref{fig:bestfit_b1b2}. For $b_1$ and $b_2$ we also plot the corresponding PBS prediction for comparison. The five different exclusion models lead to similar results. For $ R \lesssim 30 \MpcOh $, the best fit for $b_1$ agrees with the PBS prediction, especially for models 1 and 4. However, at larger void sizes, the best-fit $b_1$ turns over and starts to increase, while the PBS prediction keeps on decreasing. 
On the other hand, we find the best-fit $b_2 $ following the PBS prediction more closely, although it is smaller in magnitude at $R \gtrsim 40 \MpcOh $. We remark that in the dominant terms $P_{1,1} $ and $P_{11,11}$ only $b_1^2$ and $b_2^2$ enter, while the coupling term $P_{2,11}$ is negligible. Our fitting routine actually finds another set of acceptable fits for which  $b_2$ decreases when $R\gtrsim 30 \MpcOh $. However, even in this case  $b_1 $ increases for $R\gtrsim 30 \MpcOh $ in a way similar to what is shown in Fig.~\ref{fig:bestfit_b1b2}, mainly because it must compensate the important contribution of $P_{1,1} $ to the fit. For $D$, the best-fit values are similar to those in Fig.~\ref{fig:bestfit_b1}. Finally, with the inclusion of $b_2$, the $\chi^2  $ per degree of freedom reaches $ \sim 1 $, a big improvement compared to the linear bias model.

Recall that in Fig.~\ref{fig:bestfit_b1}, the best fit $b_1$ is systematically slightly lower than the PBS and  large-scale fit of  $b_{\rm c}$, however, as the fit is poor for the linear bias model, the results are not conclusive. With the quadratic bias, the fit is much better as manifested by $\chi^2  $ per degree of freedom is $\sim 1  $. Still, the best-fit $b_1$ is systematically lower than PBS and $b_{\rm c}$ measurements for $R \lesssim 30 \MpcOh $ and higher for larger $R$. If the  model is self-consistent we would expect that the best-fit $b_1 $ from the void-power spectrum agrees with the large-scale $b_{\rm c}$ measurement. Some of the systematics due to the construction of voids  can be eliminated by cross-correlating  dark matter with the void density field. These systematics deviate voids from being simple biased tracers of the underlying dark matter density field.  The most important one is the exclusion effect modeled by the hard-sphere model in this paper. Thus, the difference between $b_1$ from the void auto-power spectrum is likely due to  some remaining systematics in the construction of voids. This can explain why the $b_{\rm c}$ measurements agrees with the PBS results much better than those from the auto-power spectrum.     This  also suggests that  Eq.~\ref{eq:Pv_b1b2} is only a phenomenological model, it is not a serious problem if we only want to fit the void auto-power spectrum, which is often the case in galaxy surveys. We can simply marginalize over the bias parameters.

While there have already been several lines of evidence for 
corrections to the standard Poisson noise in the large-scale clustering of halos \cite{Baldaufetal2013,ChanScoccimarro2014}, our results demonstrate that this effect is even more important in the clustering of voids.

In Refs.~\cite{McDonaldRoy,ChanScoccimarroSheth2012,Baldaufetal2012} it has been proposed that at quadratic order an additional nonlocal term has to be taken into account. It was shown that this term is naturally generated by large-scale gravitational evolution \cite{ChanScoccimarroSheth2012}, and evidence for this additional term has been found in bispectrum measurements \cite{ChanScoccimarroSheth2012,Baldaufetal2012}. We will leave further investigations to future work. However, note that this nonlocal term does not affect substantially the power spectrum at large scale \cite{ChanScoccimarro2012} and, hence, is unlikely the main culprit for the lack of consistency  between  different  $b_1$ measurements.

\section{Discussion and Conclusion}
\label{sec:conclusion}

Cosmic voids have emerged as an interesting probe of cosmology, yet most studies thus far have focused on the internal characteristics of voids. 
In this paper, we have studied their large-scale clustering using $N$-body simulations and found some interesting properties. Some questions remain to be answered, we will discuss some of them below.

The voids analyzed in this paper are identified in dark matter simulation using a watershed algorithm. As the definition of voids depends on the sampling density of the tracers used to identify them, we used a few different values to investigate its effect on the size distribution, the density profile, and the power spectrum of voids. For the power spectrum we find a rather weak dependence on the sampling density of voids with radius $R \gtrsim 20 \MpcOh  $. We have cross-checked our results using sub-samples that include only parent voids, and voids whose central density is less than $ 0.2  \bar{\rho}_{\rm m} $. These different cuts on the void sample do not change the main results in this paper.   Although we only deal with dark matter voids in this paper, it is worth pointing out that besides the tracer density, the bias of tracers is also important in galaxy surveys.  We will leave it to future work to examine how the observables studied in this paper depend on the  tracer bias.

Even though the SvdW void size distribution  does not take into account either the tracer sampling-density dependence, or sub-voids, we can describe the measured void size distribution reasonable well for a large range of $R$ if the void collapse threshold is treated as a free parameter. However, the best-fit value of $\dv \sim -1  $ is very different from the canonical spherical collapse value of $-2.8$. This is very likely caused by the simplifying assumptions of the spherical expansion model. 
The spherical collapse value of $-2.8$ is obtained from the shell crossing condition, but this criterion is not incorporated in the watershed algorithm. Also,  voids constructed by  the watershed algorithm are not spherical in general and strongly depend on the sampling density.  It is also unclear why the best-fit agrees also with samples that are dominated by sub-voids (L250 with high sampling densities). One of the possible ways to investigate this further would be to trace back void particles from Eulerian to Lagrangian space 
and to study their properties in the initial density field.

We have also explored the void-matter cross-power spectrum, respectively the cross bias parameter $b_{\rm c}$ between the voids and the dark matter. We find that it can be approximated by a constant on large scales.
For our L1500 simulation, which is dominated by top-level voids, the best-fit value is broadly consistent with the PBS prediction. 
The agreement deteriorates for larger void sizes and higher redshifts. 

In order to understand the structure of $b_{\rm c}(k) $ in more detail, we have measured the void density profile in configuration space, and subsequently derived the void-matter cross-power spectrum by Fourier transform. The Fourier space oscillation in $b_{\rm c }(k)$ is clearly related to the underdense core and compensation wall seen in the void density profile. However, it is difficult to reproduce the detailed structure of $b_{\rm c}(k)$ from fitting formulae of the void density profile. 
The agreement improves for larger voids.
We have also tried to predict the void density profile by inverse Fourier transforming an educated guess of $b_{\rm c}(k)$. In this case, however, the density profile within the void is not reproduced correctly. Again, the best match is obtained for large voids. 


It is particularly important to model the void auto-power spectrum, because it is in principle directly measurable in galaxy surveys data.  Voids are very extended objects and when only parent voids are considered, they exclude each other. This has a large impact on their auto-power spectrum. We have modeled this exclusion effect using the hard-sphere model. Furthermore, to account for the fact that voids are biased tracers of the underlying dark matter density field, we have first considered the linear bias model. Combined with the hard-sphere model, we have found that it can reproduce the overall shape of the  void-auto power spectrum although  the fit is not good.  For this reason, we have extended our model to include quadratic bias using the renormalized bias approach. This improves significantly the fits, which is particularly good for voids with radii $R\gtrsim 30 \MpcOh$. 
Accounting for void exclusion is essential here because, on large scales, its sign  is opposite  that of the linear bias term while its magnitude is comparable or even larger. The shot noise renormalization term, which is constant at low $k$ and of magnitude comparable to the linear term, also contributes significantly to improve the agreement. There are some lines of evidence that this term is present in halo clustering. Void clustering statistics also support its existence.
However, the best-fit values for the linear and quadratic bias parameters $b_1$ and $b_2$ only qualitatively agree with PBS expectations. More precisely, the best-fit $b_1$ roughly agrees with the PBS prediction $b_1=1+b_1^\text{PBS}$ for $R \lesssim 30 \MpcOh $, but largely deviates from the PBS result for bigger voids. If voids would behave as dark matter halos, then $b_1$ as measured from the void auto-power spectrum should agree with $b_{\rm c}(k)$ in the limit $k\to 0$.
The lack of consistency between $b_1$ from auto-power spectrum and cross-power spectrum  could be  due to systematics in the measurements and/or modeling that have not been taken into account.

The model for the void power spectrum considered in this paper does not seem to work for the small voids.  Some insight into this problem could be gained by measuring the void auto-correlation in configuration space because it better disentangles the effect of biasing and exclusion, the latter remaining confined to short separations.

 

As already stressed several times in the literature, voids are very sensitive to the identification procedure. So are their abundance and biasing. Therefore, it is doubtful that one can ever predict accurately the abundance and clustering of the surveyed voids from first principles.   In contrast, for halos the agreement between theory and numerical results is generally more encouraging than what we find for voids. Yet, in galaxy clustering analyses, bias parameters are commonly treated as nuisance parameters.   Therefore, while theory is important for  understanding  how voids evolve in time, in practice it might be enough to devise some phenomenological description of void clustering where bias factors are marginalized over (so long as there are not too many free parameters).   Our finding that the void auto-power spectrum can be well described by a combination of exclusion and biasing is an important step, not only because it is a direct observable in galaxy surveys, but also because it enables us to infer information about the dark matter distribution in the Universe without the knowledge of the precise values of the bias parameters.


%

\section*{Acknowledgement} 
We thank Roman Scoccimarro for discussion on halo exclusion and Ravi Sheth for comments on the draft of the paper. We also thank the referee for his constructive comments.  KCC and VD are  supported by the Swiss National Science Foundation. NH is supported by the ILP LABEX (under reference ANR-10-LABX-63) and French state funds managed by the ANR within the Investissements d'Avenir program under reference ANR-11-IDEX-0004-02.  KCC acknowledges the hospitality of the Institute of Basic Science in Korea and Benasque Institute in Spain, where the last parts of this work were done. Most of the computation was performed at the cluster in the Theoretical Physics Department of the University of Geneva, while parts of the analysis were conducted at the Horizon cluster at the Institut d'Astrophysique de Paris.
\bibliography{void_exclusion} 

\end{document}